\documentclass[12pt]{article}

\usepackage[T1]{fontenc}
\usepackage[utf8]{inputenc}
\usepackage[english]{babel}
\usepackage{amsmath,amssymb,amsfonts,amsthm}            
\usepackage{mathtools,mathrsfs}                         
\usepackage{thmtools,tensor,bbm,bm}                     
\usepackage[makeroom]{cancel}                           
\usepackage{geometry,fancyhdr}                          
\usepackage{graphicx,caption,subcaption}                
\usepackage[dvipsnames,svgnames,x11names,table]{xcolor} 
\usepackage{cite}                                       
\usepackage{refcount,xr-hyper}                          
\usepackage{etoolbox,tcolorbox,array}                   
\usepackage[pdfencoding=auto, psdextra]{hyperref}       
\usepackage[capitalize]{cleveref}                       

\usepackage{faktor}
\usepackage{physics}
\usepackage{afterpage}
\usepackage{esint}

\newcommand\myshade{90}
\colorlet{mycitecolor}{JungleGreen}
\colorlet{mylinkcolor}{purple}
\colorlet{myfilecolor}{Orange2}
\colorlet{myurlcolor}{MediumSeaGreen}
\hypersetup{
    colorlinks = true,
    citecolor  = mycitecolor!\myshade!black,
    linkcolor  = mylinkcolor!\myshade!black,
    filecolor  = myfilecolor!\myshade!black,
    urlcolor   = myurlcolor!\myshade!black
}

\renewenvironment{abstract}
    {\begin{center}\bfseries Abstract.\end{center}\vspace{-30pt}
    \quotation\small\noindent\rule{\linewidth}{.5pt}\par\medskip}
    {\par\noindent\rule{\linewidth}{.5pt}\endquotation}
    \AtBeginEnvironment{abstract}{\def\quotation{\list{}{\rightmargin\leftmargin}\item\relax}\def\endquotation{\endlist}}

\newcommand{\tn}{\textnormal}
\newcommand{\GSSI}{Gran Sasso Science Institute (GSSI), I-67100 L’Aquila, Italy}
\newcommand{\GranSasso}{INFN, Laboratori Nazionali del Gran Sasso, I-67100 Assergi, Italy}

\title{\vspace{-2cm}
    \textbf{Astrophysical black holes: theory and observations}
    \vspace{0pt}}
\date{
    \today
    \vspace{-20pt}}
\author{
	Martina Adamo,$^{(a)}$\footnote{madamo@ubu.es}~~~Andrea Maselli$^{(b),(c)}$\footnote{andrea.maselli@gssi.it}\vspace{5pt}\\
    \small $^{(a)}$Departamento de F\'isica, Universidad de Burgos, 09001 Burgos, Spain \\
    \small $^{(b)}$\GSSI\\
    \small $^{(c)}$\GranSasso
    \vspace{10pt}}

\begin{document}

\maketitle
\begin{abstract}
These notes cover part of the lectures 
presented by Andrea Maselli for the 59th 
Winter School of Theoretical Physics and third 
COST Action CA18108 
Training School `Gravity -- Classical, Quantum and 
Phenomenology'. The school took place at Pałac 
Wojanów, Poland, from February 12th to 21st, 2023. 
The lectures focused on some key aspects of black hole 
physics, and in particular on the dynamics of 
particles and on the scattering of waves in the 
Schwarzschild spacetime. The goal of the course 
was to introduce the students to the concept of 
black hole quasi normal modes, to discuss their 
properties, their connection with the geodesic 
motion of massless particles, and to provide 
numerical approaches to compute their actual values.
\end{abstract}

\tableofcontents

\section{Introduction}
With nearly one hundred events from coalescing binaries 
detected by LIGO-Virgo-KAGRA \cite{LIGOScientific:2021djp}, 
gravitational-wave (GW) observations have shaped a novel 
path for studying high-energy phenomena in our Universe. 
The number of observations is expected to rise with 
current interferometers at their design sensitivity, growing 
by orders of magnitude with the next generations of ground 
and space facilities, such as the Einstein Telescope \cite{Branchesi:2023mws}, Cosmic Explorer \cite{Evans:2023euw}, and the LISA satellite \cite{LISA:2022yao}. 
The loudness of the signals detected by such network of 
GW observatories will turn GW into a new tool for 
precision (astro)physics, enabling the exploration of various scientific phenomena. Primary objectives of this quest include testing the foundations of General Relativity (GR) and understanding the nature of gravity in strong field and 
highly dynamic scenarios \cite{Berti:2015itd,Barack:2018yly,Berti:2018vdi,Berti:2018cxi}.

One of the most promising approaches for testing gravity,  
particularly a key prediction of GR, namely,
the uniqueness of Kerr black holes (BHs), revolves 
around the so-called \textit{black hole spectroscopy}. 
This approach exploits the signal following a binary 
merger, known as the ringdown, which can be described in 
terms of series of damped oscillations 
with characteristic 
frequencies called quasi-normal modes (QNMs). 
In General Relativity, QNMs are uniquely 
determined by the mass and angular momentum of the BH
\cite{Kokkotas:1999bd,Ferrari:2007dd,Berti:2009kk}. 
Measuring the frequency and damping time of a 
single QNM allows for the determination of the BH mass and spin, 
while multiple modes could provide null-hypothesis 
tests of GR \cite{Detweiler:1980gk,Dreyer:2003bv,Berti:2005ys}.
This correspondence also makes QNMs a versatile 
diagnostic tool, leading, for example, to consistency checks 
between inspiral and post-merger parameters inferred from 
binary events, searches for exotic 
states of matter at the horizon scale, 
and detection of signatures of modified theories of gravity \cite{Berti:2018vdi,Franchini:2023eda,Cardoso:2019rvt}. 
This plethora of opportunities sets the foundations 
of QNMs spectroscopy, in complete analogy with the 
longstanding efforts devoted to atomic and condensed 
matter physics. 

In these notes, we outline the essential components necessary for calculating the BH response to an external perturbation, including its QNMs spectrum. We begin by examining the fundamental properties of stationary BH solutions in GR (Section \ref{Schwarzschild sol}) and their geodesic structure (Section \ref{sec: geodesic of s}). Subsequently, our focus shifts to Section \ref{SBH}, where we delve into the formalism required to compute relativistic perturbations of Schwarzschild BHs, further explored in Section \ref{sec: perturbations}.
Throughout, we adopt geometric units, $c=G=1$.

%
\section{The Schwarzschild solution}\label{Schwarzschild sol}
%
Historically, the Schwarzschild metric represents 
the first exact solution of the Einstein 
equations, alongside Minkowski flat
spacetime, discovered by Karl Schwarzschild in 
1916, just one year after the publication of 
GR. The Schwarzschild metric is a non-trivial 
solution of the Einstein vacuum field equations
\begin{equation}\label{vacuum}
    R_{\mu\nu}=0\ ,
\end{equation}
describing a Ricci-flat manifold 
(hereafter, lowercase Greek letters represent spacetime 
indices $\mu,\nu,\dotsc=0,1,2,3$). This metric 
determines the gravitational field generated by a 
static, spherically symmetric, electrically 
uncharged, and non-rotating mass, assuming 
a vanishing cosmological constant. From a 
physical perspective, the Schwarzschild metric 
finds various applications, particularly in describing 
the vacuum outer region of non-spinning stars and planets.
%
\subsection{The Birkhoff theorem}\label{birkhoff theorem}
%
The Schwarzschild spacetime comes with a remarkable feature 
dictated by the \textit{Birkhoff theorem}. This theorem 
asserts that the Schwarzschild metric is the unique vacuum 
solution with spherical symmetry, which is also static.
In the following we shall provide the proof of the theorem. 

Consider a $(3+1)$-dimensional spacetime exhibiting spatial spherical symmetry, namely, a manifold with the three-dimensional special orthogonal group $SO(3)$ (representing rotations in three-dimensional Euclidean space) as its group of symmetries. The three generators of the action of $SO(3)$ on the spacetime are the following \cite{nakahara2018geometry}:
\begin{equation}
    \begin{aligned}
        J_1&=x^2\,\partial_3-x^3\,\partial_2=-\sin\varphi \,\partial_\theta-\cot\theta\cos\varphi \,\partial_\varphi\ ,\\
        J_2&=x^3\,\partial_1-x^1\,\partial_3=\cos\varphi \,\partial_\theta-\cot\theta\sin\varphi \,\partial_\varphi\ ,\\
        J_3&=x^1\,\partial_2-x^2\,\partial_1=\partial_\varphi\ ,
    \end{aligned}
\end{equation}
where $x^i$ are Cartesian coordinates (with spatial indices 
$i, j, \dotsc = 1, 2, 3$), and $r\in [0,+\infty)$, $\theta\in[0,\pi]$, 
$\varphi\in[0,2\pi)$ are spherical coordinates. The transformation 
between Cartesian and spherical coordinates is given by the usual 
expressions
\begin{equation}
    x^1=r \sin\theta \cos\varphi \ , \qquad x^2=r \sin\theta \sin\varphi \ , \qquad x^3=r \cos\theta \ . 
\end{equation}
The generators $J_i$ satisfy the commutation relations
\begin{equation}
    [J_i,J_j]=\varepsilon_{ijk}J_k\ ,
\end{equation}
where $\varepsilon_{ijk}$ is the Levi-Civita symbol. The 
generators of symmetries are also known as \textit{Killing 
vectors} since they satisfy the \textit{Killing equation}:
\begin{equation}\label{killing eq}
    \mathcal{L}_{J}g_{\mu\nu}=\nabla_\mu J_\nu +\nabla_\nu J_\mu =0\ ,
\end{equation}
where $\mathcal{L}_J$ is the Lie derivative along the Killing 
vector $J$ (in this case, a generator of $SO(3)$), and 
$\nabla_\mu$ represents the covariant derivative associated with 
the spacetime metric $g_{\mu\nu}$. In other words, Killing vectors 
generate transformations that preserve the metric, defining isometries.

\begin{figure}[!htb]
    \centering
    \includegraphics[width=.5\textwidth]{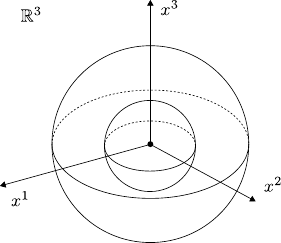}
    \captionsetup{width=.8\linewidth}
    \caption{Foliation of $\mathbb{R}^3$ (minus the origin) by two-spheres. }
    \label{fig:foliation}
\end{figure}

A three-dimensional space with $SO(3)$ as it isometry group can 
be foliated into two-spheres centered at the same origin but 
with varying radii. These two-spheres represent the homogeneous 
spaces of the $SO(3)$ group, meaning that any point on the 
sphere can be reached through a rotation starting from an 
arbitrarily chosen origin. This procedure, which 
cannot be applied to the center of the spheres, 
where the homogeneous space becomes zero-dimensional, 
is graphically depicted in Fig.~\ref{fig:foliation}. 
Each of these two-dimensional homogeneous spaces corresponds 
to a standard two-sphere with a metric, in spherical coordinates, 
given by:
\begin{equation}
    d s^2=r^2 d \Omega^2=r^2d \theta^2+r^2\sin^2\theta\, d \varphi^2\ ,
\end{equation}
where $r$ represents the radius of the sphere 
(constant within each sphere).

The process of spacetime foliation into maximally symmetric 
submanifolds, such as the two-spheres in our scenario, allows 
us to choose coordinates adapted to this foliation. For 
instance, consider a generic $n$-dimensional manifold foliated 
by $m$-dimensional submanifolds. We can use a set of $m$ 
coordinates ${u^i}$, where $i=1,\dotsc,m$, to represent the 
submanifold, and another set of $m-n$ coordinates ${v^a}$, 
where $a=1,\dotsc,m-n$, to specify the particular submanifold 
we are on. By combining these two sets of coordinates, we can 
coordinatize the entire manifold as ${u^i,v^a}$. Remarkably, 
when these submanifolds are maximally symmetric, a theorem 
(for the proof, see Ref.~\cite{weinberg1972gravitation}) 
guarantees that the metric can be expressed in the following 
form:
\begin{equation}\label{metric}
    d s^2=g_{\mu\nu}d x^\mu d x^\nu=g_{ab} (v)d v^a d v^b+f(v) h_{ij}(u) d u^i d u^j\ ,
\end{equation}
where $h_{ij}$ represents the metric of the maximally symmetric 
submanifold. We can make two important observations based on the 
form of Eq.~\eqref{metric}: (i) there are no mixed terms $d v^i d u^a$, 
namely the metric $g_{\mu\nu}$ is a block-diagonal matrix, and (ii) 
both $g_{ab}$ and $f$ depends uniquely on the variables $v^a$. The 
absence of mixed terms indicates that the submanifolds are 
consistently aligned throughout the entire space, which allows 
us to move across them while crossing points 
with the same $u^i$ coordinates but on different submanifolds. 
Additionally, the fact that $g_{ab}$ and $f$ do not depend on 
$u^i$ implies that the metric of different submanifolds remains 
the same (up to a numerical factor), as the coordinates $v^a$ 
remain constant on a given submanifold.

In our case, the submanifold coordinates are given by the 
spherical coordinates $\theta$ and $\varphi$, and the corresponding 
metric is $h_{ij}(u)d u^id u^j=d \Omega^2$. Consequently, we 
can express the metric of the entire spacetime as follows:
\begin{equation}
    d s^2=g_{11}(v)d v^1d v^1+2g_{12}(v)d v^1d v^2+g_{22}(v)d v^2d v^2+r^2(v)d \Omega^2\ ,
\end{equation}
where we have redefined the yet-to-be-determined function 
as $f(v)=r^2(v)$. To simplify the calculations, we can 
invert the function $r(v^1,v^2)$ with respect to one of 
the two variables on which it depends, for instance, 
with respect to $v^1$. Moreover, we shall find another function 
$t(v^2,r)$ such that, when expressed in terms of $t$ and $r$, 
the metric does not exhibit cross terms like $d td r$. It can 
be shown (see Ref.~\cite{carroll2019spacetime}) that this is 
always possible, which allows us to recast the metric in the 
following form:
\begin{equation}\label{metric1}
    d s^2=a_1(t,r)d t^2 + a_2(t,r) d r^2+r^2d \Omega^2\ ,
\end{equation}
The variable $r$ works as a scale factor in front of the 
metric of the two-sphere. This is also the case in the Minkowski 
spacetime in spherical coordinates, where 
$d s=-d t^2+d r^2+r^2d \Omega^2$. The latter can indeed be 
obtained by setting $a_1=-1$ and $a_2=1$ in Eq.~\eqref{metric1}, 
with the minus sign arising from the fact that the 
Minkowski spacetime is a Lorentzian manifold with 
signature $(-,+,+,+)$. Following the same procedure 
for our case, we can fix $a_1$ and $a_2$ such that
\begin{equation}\label{most generic ss metric}
    d s^2=-e^{2\alpha(t,r)} d t^2+e^{2\beta(t,r)} d r^2+r^2d \Omega^2\ .
\end{equation}
We remark that the form of Eq.~\eqref{most generic ss metric} 
is only dictated by the assumption of spherical symmetry, and 
depends on the two functions $\alpha$ and $\beta$, that 
can be determined by solving the Einstein vacuum field 
equations \eqref{vacuum}.

To solve the Einstein equations, we need to explicitly 
calculate the components of the Ricci tensor $R_{\mu\nu}$, 
derived from the Riemann curvature tensor 
$\tensor{R}{^\mu_\nu_\sigma_\rho}$:
\begin{equation}\label{eq:riccitensor}
    R_{\mu\nu}= \tensor{R}{^\sigma_\mu_\sigma_\nu}\ , \qquad \tensor{R}{^\mu_\nu_\sigma_\rho}=\partial_\sigma\Gamma^\mu{_{\rho\nu}}-\partial_\rho\Gamma^\mu{_{\sigma\nu}}+\Gamma^\mu{_{\sigma\lambda}}\Gamma^\lambda{_{\rho\nu}}-\Gamma^\mu{_{\rho\lambda}}\Gamma^\lambda{_{\sigma\nu}}\ ,
\end{equation}
where $\Gamma^\mu{_{\nu\sigma}}$ are the Christoffel 
symbols
\begin{equation}\label{christoffel symbols}
    \Gamma^\mu{_{\nu\sigma}}=\frac{1}{2}g^{\mu\lambda}(\partial_\nu g_{\lambda \sigma}+\partial_\sigma g_{\lambda \nu}- \partial_\lambda g_{\nu\sigma})\ .
\end{equation}
In our case, the only non-zero components of the metric are 
given by the diagonal terms $g_{tt}=-e^{2\alpha(t,r)}$, $g_{rr}=e^{2\beta(t,r)}$, $g_{\theta\theta}=r^2$, $g_{\varphi\varphi}=r^2\sin^2\theta$. Replacing the 
former into Eq.~\eqref{eq:riccitensor}, we obtain:
\begin{equation}
    \begin{aligned}
        R_{tt}&=\partial_t^2\beta+(\partial_t\beta)^2- \partial_t\alpha\,\partial_t\beta+e^{2(\alpha-\beta)}\left[\partial_r^2\alpha+(\partial_r\alpha)^2- \partial_r\alpha\,\partial_r\beta + \frac{2}{r} \partial_r\alpha \right]\ ,\\
        R_{rr}&=- \partial_r^2\alpha-(\partial_r\alpha)^2+ \partial_r\alpha\,\partial_r\beta + \frac{2}{r} \partial_r\alpha 
        + e^{-2(\alpha-\beta)}\left[\partial_t^2\beta
        +(\partial_t\beta)^2- \partial_t\alpha\,\partial_t\beta \right]\ ,\\
        R_{tr}&=R_{rt}=\frac{2}{r}\partial_t \beta\ ,\\
        R_{\theta\theta}&=1+e^{-2\beta}\left(r\partial_r\beta-r\partial_r\alpha-1\right)\ ,\\
        R_{\varphi\varphi}&=R_{\theta\theta}\sin^2\theta\ ,
    \end{aligned}
\end{equation}
with all the other components vanishing. We require  
each term to be zero. The simplest to solve is given 
by $R_{tr}=0$, which tells us that the function $\beta$ 
depends only on $r$. This provides a significant simplification, 
as time derivatives can be set to zero in all the other 
Ricci components. 
Moreover, we can differentiate $R_{\theta\theta}$ with respect to 
$t$, yielding $\partial_t\partial_r \alpha=0$. This means that the 
function $\alpha$ can be expressed as the sum of a function depending 
solely on $r$ and another depending only on $t$, namely $\alpha(t,r)=\alpha_1(r)+\alpha_2(t)$. With these results, we can 
rewrite the metric as follows:
\begin{equation}
    d s^2=-e^{2\alpha_1(r)}e^{2\alpha_2(t)} d t^2+e^{2\beta(r)} d r^2+r^2d \Omega^2\ .
\end{equation}
However, we can always change the variable $t$ to a new time 
coordinate $t'$ such that $d t'=e^{\alpha_2(t)}d t$ and, in 
terms of $t'$, the metric reads
\begin{equation}\label{static metric}
    d s^2=-e^{2\alpha_1(r)} d t'^2+e^{2\beta(r)} d r^2+r^2d \Omega^2\ .
\end{equation}
We relabel for sake of simplicity $\alpha_1$ and $t'$ as 
$\alpha$ and $t$, obtaining a metric that in these coordinates 
does not depend explicitly on time. This is a key result of 
the Birkhoff theorem, \textit{i.e.}, a spherically symmetric 
gravitational field in empty space must be static.  

Before proceeding with our calculations, let us remind 
that a metric is said to be \textit{stationary} if 
it appears the same at each instant of time, implying the 
existence of a timelike Killing vector. By choosing 
coordinates adapted to this Killing vector, the metric 
does not depend on time. The most general stationary 
metric can be written as
\begin{equation}
    d s^2=g_{00}(\vec{x})d x^0 d x^0+2g_{0i}(\vec{x}) d x^0 d x^i+g_{ij}(\vec{x})d x^i d x^j\ .
\end{equation}
If we further ask the metric to be static,  in addition to requiring the existence of a timelike Killing vector, we also require this vector to be orthogonal to a family of spacelike hypersurfaces. 
This condition leads to the absence of cross terms $d t d x^i$ in the 
metric, namely:
\begin{equation}\label{eq:metricstaticstation}
    d s^2=g_{00}(\vec{x})d x^0 d x^0+g_{ij}(\vec{x})d x^i d x^j\ .
\end{equation}
The spherically symmetric solution we derived in 
Eq.~\eqref{static metric}, expressed in the coordinates 
$r$ and $t$ with respect to which it is time-independent, 
is in the form of Eq.~\eqref{eq:metricstaticstation}.

We are then left with the following set of equations to solve:
\begin{equation}
    \begin{aligned}
        R_{tt}&=e^{2(\alpha-\beta)}\left[\partial_r^2\alpha+(\partial_r\alpha)^2- \partial_r\alpha\,\partial_r\beta + \frac{2}{r} \partial_r\alpha \right]=0\ ,\\
        R_{rr}&=- \partial_r^2\alpha-(\partial_r\alpha)^2+ \partial_r\alpha\,\partial_r\beta + \frac{2}{r} \partial_r\alpha =0\ ,\\
        R_{\theta\theta}&=1+e^{-2\beta}(r\partial_r\beta-r\partial_r\alpha-1)=0\ ,\\
        R_{\varphi\varphi}&=R_{\theta\theta}\sin^2\theta=0\ ,
    \end{aligned}
\end{equation}
An interesting observation is that if $R_{\theta\theta}=0$, it 
automatically implies $R_{\varphi\varphi}=0$, so we do not need to worry 
about the latter. Moreover, since $R_{tt}$ and $R_{rr}$ must vanish 
independently, this condition also applies to their linear 
combination
\begin{equation}
    e^{-2(\alpha-\beta)}R_{tt}+R_{rr}=\frac{2}{r}\partial_r(\alpha+\beta)=0\ ,
\end{equation}
implying $\alpha+\beta$ is a constant, or equivalently 
$\alpha=-\beta+(constant)$. However, we can rescale the time 
coordinate $t\to t'=e^{(constant)}t$ to reabsorb this factor, 
leading the metric to read (once again, relabeling $t'$ as $t$):
\begin{equation}
    d s^2=-e^{-2\beta(r)} d t^2+e^{2\beta(r)} d r^2+r^2d \Omega^2\ .
\end{equation}
Focusing now on $R_{\theta\theta}=0$, and using the expression 
$\alpha=-\beta$ we obtained above, the equation becomes
\begin{equation}
    R_{\theta\theta}=1+e^{-2\beta}(r\partial_r\beta-r\partial_r\alpha-1)=1-e^{-2\beta}(1-2r\partial_r\beta)=0\ ,
\end{equation}
which simplifies to
\begin{equation}
    \partial_r(r e^{-2\beta})=1\ ,
\end{equation}
The solution is given by $e^{-2\beta}=1-R_S/r$, where $R_S$ 
is an integration constant. Direct calculations confirm 
that this expression for $e^{-2\beta}$ satisfies $R_{tt}=R_{rr}=0$. 
As a result, the metric takes the form
\begin{equation}\label{Schwarzschild metric}
    d s^2=-\left( 1-\frac{R_S}{r} \right) d t^2+\left( 1-\frac{R_S}{r} \right)^{-1} d r^2+r^2d \Omega^2\ .
\end{equation}
This is the Schwarzschild metric, obtained as a solution of the vacuum 
Einstein equations, assuming spherical symmetry. 
%
\subsection{Physical interpretation of the Schwarzschild radius}\label{Schar}
%
The Schwarzschild metric we derived is actually a one-parameter 
family of solutions, which depends on the \textit{Schwarzschild radius} 
$R_S$. This parameter holds a straightforward physical interpretation 
in the weak-field regime. In such regime, we can consider a curved 
metric as a small perturbation of the Minkowski flat spacetime 
$\eta_{\mu\nu}=\operatorname{diag}(-1,1,1,1)$:
\begin{equation}\label{weak}
    g_{\mu\nu}=\eta_{\mu\nu}+h_{\mu\nu}\ , \qquad |h_{\mu\nu}|\ll 1\ .
\end{equation}
We focus moreover on the Newtonian limit, such that test 
particles move slowly in a weak and stationary gravitational 
field. This implies that particles are non-relativistic, 
and their four-velocity components satisfy
\begin{equation}\label{nra}
    \frac{d x^i}{d \tau} \ll \frac{d x^0}{d \tau}\ ,
\end{equation}
where $\tau$ is the proper time. The motion of the particle is 
then described by the geodesic equation, which in this case 
simplifies as follows:
\begin{equation}
    \frac{d^2 x^\mu}{d \tau^2}+\Gamma^\mu{_{\nu\sigma}}  \frac{d x^\nu}{d \tau} \frac{d x^\sigma}{d \tau}\simeq  \frac{d^2 x^\mu}{d \tau^2}+\Gamma^\mu{_{00}}  \frac{d x^0}{d \tau} \frac{d x^0}{d \tau}=0\ ,
\end{equation}
where all terms involving $\frac{d x^i}{d \tau}$ have been 
neglected due to the non-relativistic assumption \eqref{nra}. 
Furthermore, since the gravitational field is stationary, all 
time derivatives of the metric vanish, and the Christoffel 
symbols become:
\begin{equation}
    \Gamma^\mu{_{00}}=-\frac{1}{2}g^{\mu\nu}\partial_\nu g_{00}\simeq -\frac{1}{2}\eta^{\mu\nu}\partial_\nu h_{00}\ ,
\end{equation}
where we used of the weak-field \textit{ansatz} \eqref{weak}. As 
a result, the time component of the geodesic equation 
simply becomes $\frac{d x^0}{d \tau}= \text{constant}$. 
For the space components, we find
\begin{equation}
    \frac{d^2 x^i}{d \tau^2}-\frac{1}{2}\delta^{ij}\partial_j h_{00}  \frac{d x^0}{d \tau} \frac{d x^0}{d \tau}=0\ ,
\end{equation}
which can be rewritten as
\begin{equation}
    \frac{d^2 \vec{x}}{d \tau^2} \frac{d \tau}{d x^0}\frac{d \tau}{d x^0}=\frac{d^2 \vec{x}}{(d x^0)^2}=\frac{1}{2}\vec{\nabla}h_{00}\ .
\end{equation}
When we compare this with the corresponding Newtonian 
equation
\begin{equation}
    \frac{d^2 \vec{x}}{(d x^0)^2}=-\vec{\nabla} \Phi\ ,\label{eq:newtonf}
\end{equation}
which describes the acceleration of a particle in a 
gravitational potential $\Phi=-\frac{M}{r}$ generated 
by a mass $M$ at a distance $r$ from the particle.  
Requiring that in the Newtonian limit we recover the classical 
result, \eqref{eq:newtonf} implies that
\begin{equation}
    h_{00}=-2\Phi+(constant)\ .
\end{equation}
Moreover, asking the metric to approach the Minkowskian solution 
at spatial infinity from the gravitational source, leads the 
integration constant to vanish, such that the perturbation 
$h_{00}$ reads
\begin{equation}\label{weakmetric}
    h_{00}=-2\Phi\ ,\qquad \Rightarrow \qquad g_{00}=-(1+2\Phi)=-\left(1-\frac{2M}{r}\right)\ .
\end{equation}
We can apply this arguments to the Schwarzschild metric. Far 
from the source, for $r\gg R_S$, in the weak-field regime, 
the solution must approach the form of Eq.~\eqref{weakmetric}. 
This allows us to identify $R_S=2M$, where the mass $M$ 
is the source of the gravitational field. In the limit 
where the ratio $M/r$ is small, \textit{i.e}, 
when $M\to 0$ or $r\to \infty$, we recover the Minkowskian 
spacetime, such that the metric is \textit{asymptotically flat}.
%
\subsection{The Schwarzschild singularity} 
%
Studying the Schwarzschild metric \eqref{Schwarzschild metric}, 
we can identify two problematic values of the radial coordinate, namely $r=0$ and $r=R_S$. In both cases, one of the metric components vanishes while another tends to infinity. However, such metric components depend on the choice of coordinates. 
This poses the problem of determining whether the values 
$r=0$ and $r=R_S$ correspond to physical singularities or if they 
are artifacts given by our particular coordinate system. 
To address this issue, we need to study quantities that 
characterize the curvature of the manifold in a coordinate-independent way. These quantities are scalars constructed 
from the Riemann curvature tensor and the metric.

In a $N$-dimensional manifold, the Riemann tensor and the 
metric have $\frac{1}{12}N^2(N^2-1)$ and $\frac{1}{2}N(N+1)$ independent components, respectively. However, with a change 
of variables we can locally fix $N^2$ of them. As a result, the number of independent scalars that can be constructed from 
$R_{\alpha\beta\gamma\delta}$ and $g_{\alpha\beta}$ is 
given by
\begin{equation}
    \frac{1}{12}N^2(N^2-1)+\frac{1}{2}N(N+1)-N^2=\frac{1}{12}N(N-1)(N-2)(N+3)\ .
\end{equation}
Note that for $N=1,2$ the previous equation predicts zero 
curvature invariants. However, in two dimensions (which is 
the only exception for this argument), we do have one 
curvature invariant, namely the Ricci scalar. In four 4D 
we have 14 curvature invariants, which can be enumerated using the following decomposition of the Riemann tensor:
\begin{equation}\label{Riemann decomposition}
    \begin{aligned}
        R_{\mu\nu\sigma\rho}&= \frac{1}{N-2}(g_{\mu\sigma}R_{\nu\rho}+g_{\nu\rho}R_{\mu\sigma}-g_{\mu\rho}R_{\nu\sigma}-g_{\nu\sigma}R_{\mu\rho})\\
        & \qquad\qquad-\frac{1}{(N-1)(N-2)}(g_{\mu\sigma}g_{\nu\rho}-g_{\mu\rho}g_{\nu\sigma})R+C_{\mu\nu\sigma\rho}\ ,
    \end{aligned}
\end{equation}
where $C_{\mu\nu\sigma\rho}$ is the \textit{Weyl tensor}, \textit{i.e.}, 
the traceless part of the Riemann tensor. The Weyl tensor is related to conformal deformations of spacetime. Like the Riemann tensor, it 
measures spacetime curvature, but it retains only information about 
shape deformations, while does not take into account changes in 
volume.

Looking at the decomposition \eqref{Riemann decomposition} we 
immediately observe that in Ricci-flat manifolds, such that 
$R_{\mu\nu}=0$ and $R=0$, the Weyl tensor provides the only non-zero 
component of the Riemann tensor. If the Weyl tensor is also zero, this implies that the metric is conformally flat. 
Within Ricci-flat manifolds, 10 of the 14 curvature invariants are 
given by $R_{\mu\nu}=0$, which represents an invariant statement 
despite the Ricci tensor not being a scalar. The remaining four 
curvature invariants are given by
\begin{equation}
    \begin{array}{ll}
       C^{\mu\nu\sigma\rho}C_{\mu\nu\sigma\rho}\ , & \frac{1}{\sqrt{g}}\tensor{\varepsilon}{^\mu^\nu_\lambda_\tau}C^{\lambda\tau\sigma\rho}C_{\mu\nu\sigma\rho}\ , \\ \\
       C_{\mu\nu\sigma\rho}C^{\sigma\rho\lambda\tau}\tensor{C}{_\lambda_\tau^\mu^\nu}\ , \qquad  & \frac{1}{\sqrt{g}} C_{\mu\nu\sigma\rho}C^{\sigma\rho\lambda\tau}\tensor{\varepsilon}{_\lambda_\tau^\xi^\kappa}\tensor{C}{_\xi_\kappa^\mu^\nu}\ .
    \end{array}
\end{equation}
These expressions allow us to compute the curvature invariants 
for the Schwarzschild metric \eqref{Schwarzschild metric}. 
Coming back to our problem, the invariants we defined 
are finite when evaluated at $r=R_S$, while become singular for 
$r=0$. This shows that the ``singularity'' at the Schwarzschild 
radius is merely a coordinate singularity, and it is possible to 
identify a coordinate system where the metric is well-behaved at 
$R_S$. On the other hand, the singularity at the origin is 
genuine and retains its character regardless of the coordinate 
system \cite{weinberg1972gravitation}.

However, it is important to reiterate that the Schwarzschild metric we have derived applies exclusively in vacuum: it remains valid only \textit{outside} the massive spherical body, that is, the source of the metric, such as a planet or a star. For example, if we consider the Sun with a radius of $R_\odot=10^6 M_\odot$, significantly surpassing its Schwarzschild radius $R_{S\,\odot}=2M_\odot$, we find both the Schwarzschild radius and the origin of coordinates to be inside the Sun, which is however described by a different metric and the Schwarzschild solution does not apply anymore. Nevertheless, there exist objects like black holes for which the exterior metric is valid everywhere, as we will see in Sec.~\ref{SBH} \cite{carroll2019spacetime}.

\section{Geodesics of Schwarzschild} \label{sec: geodesic of s}
%
In this section, our focus is on the geodesic structure of the Schwarzschild spacetime, providing a clear physical interpretation of the QNM frequencies discussed in the subsequent sections. As briefly mentioned in Sec.~\ref{Schar}, geodesics represent the paths followed by free particles in a given spacetime, and their trajectory is described by the equation
\begin{equation}\label{geodesic eqs}
    \frac{d^2 x^\mu}{d \tau^2}+\Gamma^\mu_{\nu\sigma} \frac{d x^\nu}{d \tau} \frac{d x^\sigma}{d \tau}=0\ ,
\end{equation}
where the the Christoffel symbols $\Gamma^\mu_{\nu\sigma}$ 
are defined in Eq.~\eqref{christoffel symbols}. 
Geodesics are curves that parallel transport their own 
tangent vector. Representing the ``strightest'' path on a manifold, 
they provide a local extremum for the length of a curve 
connecting two points. Indeed, the geodesic equation 
\eqref{geodesic eqs} can be derived from a variational 
principle, starting with the action of a free test 
particle
\begin{equation}\label{free particle action}
    S[x]=\int\! d \lambda\, \sqrt{g_{\mu\nu}(x)\frac{d x^\mu}{d \lambda}\frac{d x^\nu}{d \lambda}}\ ,
\end{equation}
where $\lambda$ is the curve parameter. Variation of $S[x]$ 
yields the Euler--Lagrange equations of motion
\begin{equation}
    \frac{\partial L}{\partial x^\mu}-\frac{d}{d \lambda} \frac{\partial L}{\partial \dot x^\mu}=0\ ,
\end{equation}
where dots identify derivatives with respect to the parameter 
$\lambda$, and the Lagrangian $L(x,\dot x, \lambda)$ is given by
\begin{equation}
    L=\sqrt{g_{\mu\nu}(x)\frac{d x^\mu}{d \lambda}\frac{d x^\nu}{d \lambda}} \ .
\end{equation}
Calculations of the Euler--Lagrange equations leads to
\begin{equation}
    \ddot x^\mu + \frac{1}{2} g^{\mu\nu}(\partial_\sigma g_{\nu\rho}+\partial_\rho g_{\nu\sigma}-\partial_\nu g_{\sigma\rho})\dot x^\sigma \dot x^\rho=0\ .
\end{equation}
Using the definition of Christoffel symbols discussed in the previous sections (see Eq.~\eqref{christoffel symbols}), this calculation immediately leads to the geodesic equation \eqref{geodesic eqs}. However, geodesic equations can be derived in various ways, including a direct application of the equivalence principle \cite{weinberg1972gravitation}.
%
\subsection{Constants of motion}
%
Using the explicit expressions of the Christoffel symbols for 
the Schwarzschild metric
\begin{equation}
    \begin{aligned}
        \Gamma^r{_{rr}}&=-\tfrac{M}{r(r-2M)}\ ,\\
        \Gamma^r{_{\varphi\varphi}}&=-(r-2M)\sin^2\theta\ ,\\
        \Gamma^\theta{_{\varphi\varphi}}&=-\sin\theta\cos\theta\ ,
    \end{aligned}
    \qquad
    \begin{aligned}
        \Gamma^r{_{tt}}&=\tfrac{M(r-2M)}{r^3}\ ,\\
        \Gamma^t{_{tr}}&=\tfrac{M}{r(r-2M)}\ ,\\
        \Gamma^\varphi{_{r\varphi}}&=\tfrac{1}{r}\ ,
    \end{aligned}
    \qquad
    \begin{aligned}
        \Gamma^r{_{\theta\theta}}&=-(r-2M)\ ,\\
        \Gamma^\theta{_{r\theta}}&=\tfrac{1}{r}\ ,\\
        \Gamma^\varphi{_{\theta\varphi}}&=\tfrac{\cos\theta}{\sin\theta}\ ,
    \end{aligned}
\end{equation}
we can derive the four components of the geodesic 
equation
\begin{equation}
    \begin{aligned}
        \tfrac{d^2 t}{d \lambda}^2&=-\tfrac{2M}{r(r-2M)}\tfrac{d r}{d \lambda}\tfrac{d t}{d \lambda}\ ,\\
        \tfrac{d^2 r}{d \lambda}^2&=-\tfrac{M(r-2M)}{r^3}\left(\tfrac{d t}{d \lambda}\right)^2+\tfrac{M}{r(r-2M)}\left(\tfrac{d r}{d \lambda}\right)^2+(r-2M)\left[\left(\tfrac{d \theta}{d \lambda}\right)^2+\sin^2\theta\left(\tfrac{d \varphi}{d \lambda}\right)^2\right]\ ,\\
        \tfrac{d^2 \theta}{d \lambda}^2&=-\tfrac{2}{r} \tfrac{d \theta}{d \lambda}\tfrac{d r}{d \lambda}+\sin\theta\cos\theta \left(\tfrac{d \varphi}{d \lambda}\right)^2\ ,\\ 
        \tfrac{d^2 \varphi}{d \lambda}^2&=-\tfrac{2}{r}\tfrac{d r}{d \lambda}\tfrac{d \varphi}{d \lambda}-\tfrac{2\cos\theta}{\sin\theta}\tfrac{d \theta}{d \lambda}\tfrac{d \varphi}{d \lambda}\ .
    \end{aligned}
\end{equation}
These equations form a system of coupled ordinary differential 
equations that can be solved by taking advantage of the 
symmetries of the Schwarzschild metric. As discussed in 
Sec.~\ref{birkhoff theorem}, the spherically symmetric 
Schwarzschild metric possesses three Killing vectors (the generators of the action of $SO(3)$ on spacetime). Furthermore, 
the Birkhoff theorem establishes that the unique vacuum solution 
with spherical symmetry must also be static, implying the existence 
of a timelike Killing vector. Consequently, the Schwarzschild 
metric possesses four Killing vectors:
\begin{equation}
    \begin{aligned}
        J_0&=\partial_t\ ,\\
        J_1&=-\sin\varphi \,\partial_\theta-\cot\theta\cos\varphi \,\partial_\varphi\ ,\\
        J_2&=\cos\varphi \,\partial_\theta-\cot\theta\sin\varphi \,\partial_\varphi\ ,\\
        J_3&=\partial_\varphi\ ,
    \end{aligned}
\end{equation}
where $J_0$ represents the timelike Killing vector, orthogonal to spacelike hypersurfaces, generating time translations. Meanwhile, $J_{i=1,2,3}$ are the three Killing vectors associated with spatial rotations. The geodesic equation \eqref{geodesic eqs} can be rewritten in the compact form as follows:
\begin{equation}
    U^\mu \nabla_\mu U^\nu=0\ ,\label{eq:geodesic2}
\end{equation}
where $U^\mu=\dot x^\mu$ is the tangent vector. It is straightforward to prove that $U^\mu J_\mu$ is a constant of motion associated with the Killing vector $J$. Indeed,
\begin{equation}
    U^\mu\nabla_\mu(U^\nu J_\nu)=U^\mu\nabla_\mu U^\nu J_\nu+U^\mu U^\nu \nabla_\mu J_\nu=U^\mu\nabla_\mu U^\nu J_\nu+\frac{1}{2}U^\mu U^\nu (\nabla_\mu J_\nu+\nabla_\nu J_\mu)=0\ .
\end{equation}
This expression vanishes since both terms in the last equality are zero, due to Eq.~\eqref{eq:geodesic2} and the Killing equation, respectively. Therefore, we have four conserved quantities.

Since the Schwarzschild metric is asymptotically flat, we can 
determine the physical meaning of these quantities studying 
their far-field limit, \textit{i.e.}, their behavior at 
large spatial distances. The constant of motion associated 
with the invariance under time translations can be interpreted 
as the energy per unit mass of the particle. Constants related 
to the generators of spatial rotations can be interpreted as 
the three components of angular momentum. For the Schwarzschild 
metric in particular we have for $J_0$ and $J_3$:
\begin{equation}\label{conserved quantities}
    \begin{aligned}
        E&=-g_{\mu\nu}U^\mu J_0^\nu=\left(1-\frac{2M}{r}\right)\frac{d t}{d \lambda}\ ,\\
        L&=g_{\mu\nu}U^\mu J_3^\nu=r^2 \sin^2\theta\,\frac{d \varphi}{d \lambda}=r^2 \frac{d \varphi}{d \lambda}\ ,
    \end{aligned}
\end{equation}
where $E$ and $L$ are the energy of the test particle, and the magnitude of its angular momentum. Note that if the 
direction of the latter is conserved, the motion is constrained 
to a fixed plane during time evolution. We always have the 
freedom to rotate our coordinate system in such a way that 
this plane coincides with the equatorial one, \textit{i.e.}, 
to set $\theta=\pi/2$.

Finally, it is worth mentioning that another constant of 
motion exists, which can be derived directly recognising 
that the metric itself is a trivial solution of the 
Killing equation, being the connection compatible with 
the metric, $\nabla_\mu g_{\nu\sigma}=0$. Therefore
\begin{equation}
    U^\mu\nabla_\mu(g_{\nu\sigma}U^\nu U^\sigma)=U^\mu\nabla_\mu g_{\nu\sigma} U^\nu U^\sigma + 2 g_{\nu\sigma} U^\nu U^\mu\nabla_\mu U^\sigma =0\ ,\label{eq:consU}
\end{equation}
namely $U^\mu U_\mu$ is a constant motion, with  $U^\mu U_\mu<0$ 
and $U^\mu U_\mu=0$ for massive and massless particles, 
respectively. Note also that for massive bodies we can choose 
the proper time\footnote{Although this is not the case for 
massless particles,  it is always possible to find an affine 
parameter such that the geodesic equation for massless particles 
is given by Eq.~\eqref{geodesic eqs}.} $\tau$ to parametrize 
the geodesic, such that $U^\mu U_\mu=-1$. 

In the case of the Schwarzschild spacetime, Eq.~\eqref{eq:consU} 
takes the explicit form
\begin{equation}\label{character geodesic}
    U^\mu U_\mu= g_{\mu\nu} \frac{d x^\mu}{d \lambda} \frac{d x^\nu}{d \lambda}= -\left(1-\frac{2M}{r}\right) \left(\frac{d t}{d \lambda}\right)^2 + \left(1-\frac{2M}{r}\right)^{-1} \left(\frac{d r}{d \lambda}\right)^2+r^2 \left(\frac{d \varphi}{d \lambda}\right)^2=\epsilon\ ,
\end{equation}
where $\epsilon=-1,0$ for a massive and massless test 
particles, respectively. Using Eq.~\eqref{conserved quantities} we can rewrite 
Eq.~\eqref{character geodesic} as
\begin{equation}
    -E^2+\left(\frac{d r}{d \lambda}\right)^2+\left(1-\frac{2M}{r}\right)\left(\frac{L^2}{r^2}-\epsilon\right)=0\ ,
\end{equation}
which is a differential equation for the variable $r(\lambda)$. 
This equation can be recast in the following form:
\begin{equation}
    \frac{1}{2} \left(\frac{d r}{d \lambda}\right)^2+ V_{e\!f\!f}(r)=\mathcal{E}\ ,
\end{equation}
where $V_{e\!f\!f}(r)$ is the radial-dependent effective 
potential given by
\begin{equation}\label{Veff}
    V_{e\!f\!f}(r)=-\frac{1}{2}\epsilon+\epsilon\frac{M}{r}+\frac{L^2}{2r^2} -\gamma \frac{ML^2}{r^3}\ , \qquad \mathcal{E}=\frac{1}{2}E^2\ .
\end{equation}
The second term on the right-hand side represents the standard gravitational potential, while the third and fourth components account for angular momentum contributions. The parameter $\gamma$ allows a direct comparison between General Relativity (GR) and Newtonian Gravity (NG), namely, $\gamma=0,1$ in NG and GR, respectively \cite{carroll2019spacetime}.
The effective potential takes the form of a $1/r$ power series, 
which makes different terms being more or less relevant at 
different scales. In particular, at large distances, the 
Newtonian and GR descriptions align, while for small values 
of $r$, the relativistic contribution induced by angular 
momentum becomes more relevant.

Summarizing, the effective potential provided by Eq.~\eqref{Veff} 
allows us to study the orbits of both massive particles 
($\epsilon=-1$) and massless particles ($\epsilon=0$) moving 
in the gravitational field produced by a mass $M$ located at 
the origin of the coordinates, in both GR ($\gamma=1$) and 
NG ($\gamma=0$). 
Our analysis uses Schwarzschild coordinates, which makes 
problematic to describe geodesics at $R_S=2M$, and requires 
the introduction of non-singular coordinates, which we 
discuss in Sec.~\ref{SBH}. We will now focus in details on the 
features of the geodesics for the two values of $\epsilon$ we 
considered.
%
\subsection{Orbits of massive particles}
%
For massive particles, $\epsilon=-1$, the effective 
potential \eqref{Veff} is given by
\begin{equation}
    V_{e\!f\!f}(r)=\frac{1}{2}-\frac{M}{r}+\frac{L^2}{2r^2} -\gamma \frac{ML^2}{r^3}\ .
\end{equation}
Our goal is to study the behavior of this function within 
GR and NG. As $r$ approaches 
$+\infty$, the potential tends to the same 
limit $V_{e\!f\!f}(r)\to\frac{1}{2}$. On the other side of 
the domain, for $r\rightarrow 0$, the form of the potential 
depends on $\gamma$:
\begin{equation}\label{limitbehavior}
    \lim_{r\to 0} V_{e\!f\!f}(r)=\begin{cases}
        +\infty\ , \qquad\gamma=0\ , \ \ (\text{NG})\ ,\\
        -\infty\ , \qquad\gamma=1\ , \ \ (\text{GR})\ .\\
    \end{cases}
\end{equation}
By searching for extrema of the effective potential, 
we find two roots
\begin{equation}\label{eq:rootsVeff}
    r_{\pm}=\frac{L^2\pm \sqrt{L^4-12\gamma (ML)^2}}{2M}\ ,
\end{equation}
which are particularly useful for distinguishing between 
the GR and NG scenarios.

\begin{figure}[!htb]
    \centering
    \includegraphics[width=0.69\textwidth]{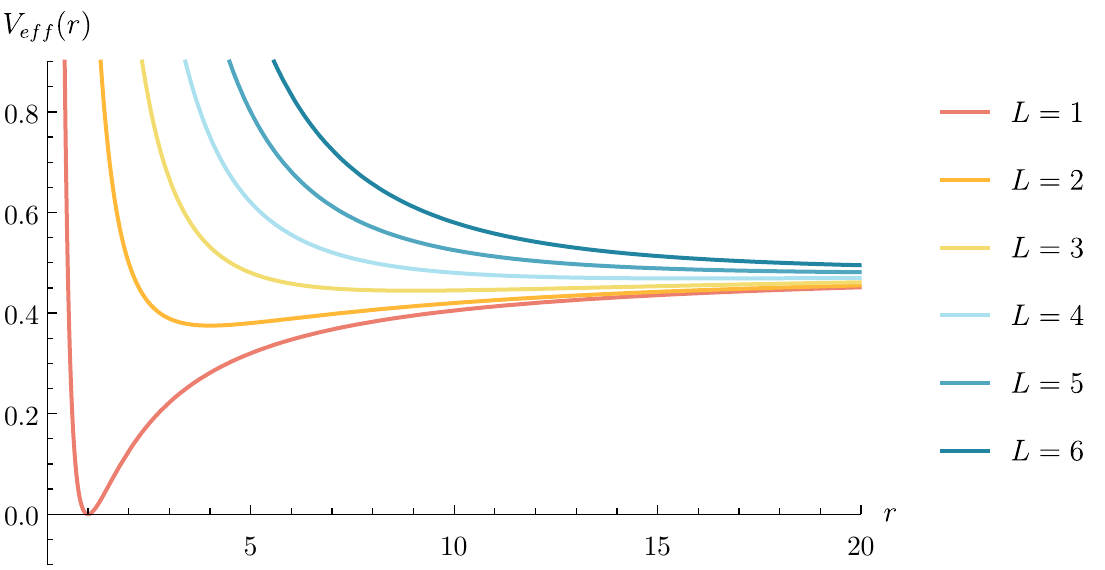}
    \captionsetup{width=.8\linewidth}
    \caption{Shape of the effective 
    potential $V_{e\!f\!f}$ as a function of the 
    coordinate radius, for massive particles in 
    Newtonian gravity. Colored curves refer to different 
    values of $L$ (we scale lengths such that $M=1$). 
    As the angular momentum increases, the radius $r_*$ 
    corresponding to the the minimum of the potential 
    also increases, and so does the value of the 
    potential $V_{e\!f\!f}(r_*)$. Note that the 
    potential tends to $1/2$ as $L\to+\infty$.}
    \label{fig:NGmassive}
\end{figure}

\begin{figure}[!htb]
    \centering
    \includegraphics[width=0.69\textwidth]{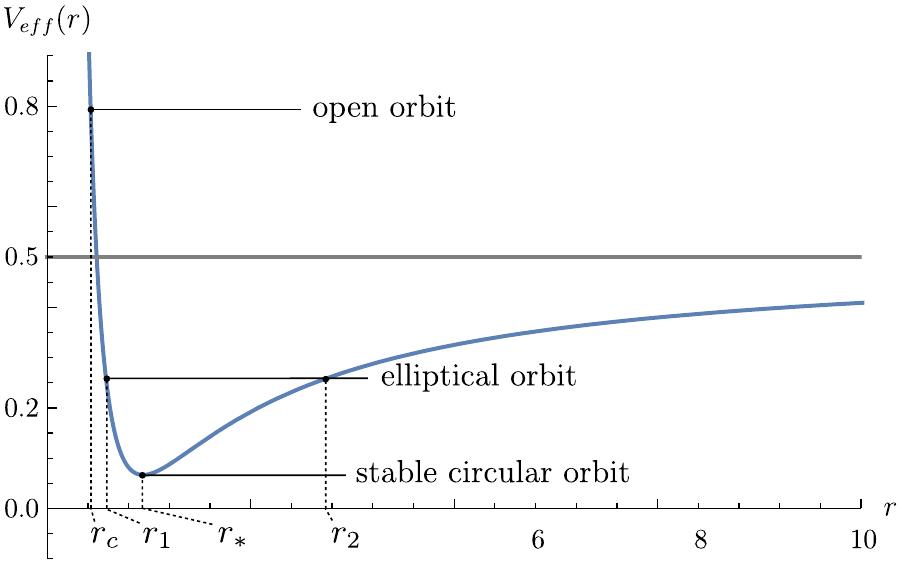}
    \captionsetup{width=.8\linewidth}
    \caption{Effective potential $V_{e\!f\!f}$ 
    (blue curve) as a function of the radial coordinate $r$ 
    for massive particles in Newtonian gravity, assuming 
    $L=1.07$ ($M=1$). For large values of $r$, the effective 
    potential approaches the asymptotic limit $V_{eff}\rightarrow\frac{1}{2}$ (gray horizontal line). 
    We show as black lines three possible cases for the 
    energy $\mathcal{E}$: (i) if the energy matches the 
    minimum of the potential, the particle remains in a 
    stable circular orbit with radius $r_*$; (ii) if $\mathcal{E}$ 
    falls between the asymptotic limit and the minimum, 
    the orbit becomes elliptical, with the radius oscillating 
    between $r_1$ and $r_2$; (iii) for energies higher than 
    the asymptotic value, the particle approaches the source up 
    to a minimum radius $r_c$ and then moves on an open orbit.}
    \label{fig:NGmassive2}
\end{figure}

\paragraph{Massive particles in Newtonian gravity.} For 
$\gamma=0$ the two roots of Eq.~\eqref{eq:rootsVeff} 
collapse into a single value, $r_*=L^2/M$. The behavior 
of the effective potential in this case is shown in 
Figs.~\ref{fig:NGmassive}-\ref{fig:NGmassive2} as a 
function of $r$, for different choices of $L$. We recognise 
three regimes. If the test particle approaches the source 
with an energy $\mathcal{E}$ equal to $V_{e\!f\!f}(r_*)$, 
it will remain bound in a stable circular orbit with radius 
$r_*$. When $V_{e\!f\!f}(r_*)<\mathcal{E}<\frac{1}{2}$, 
the orbit becomes elliptic, swinging around the radius 
of the stable circular orbit. The third scenario occurs 
when $\mathcal{E}\geq \frac{1}{2}$, and the particle 
follows an open orbit.

\begin{figure}[!htb]
    \centering
    \includegraphics[width=0.7\textwidth]{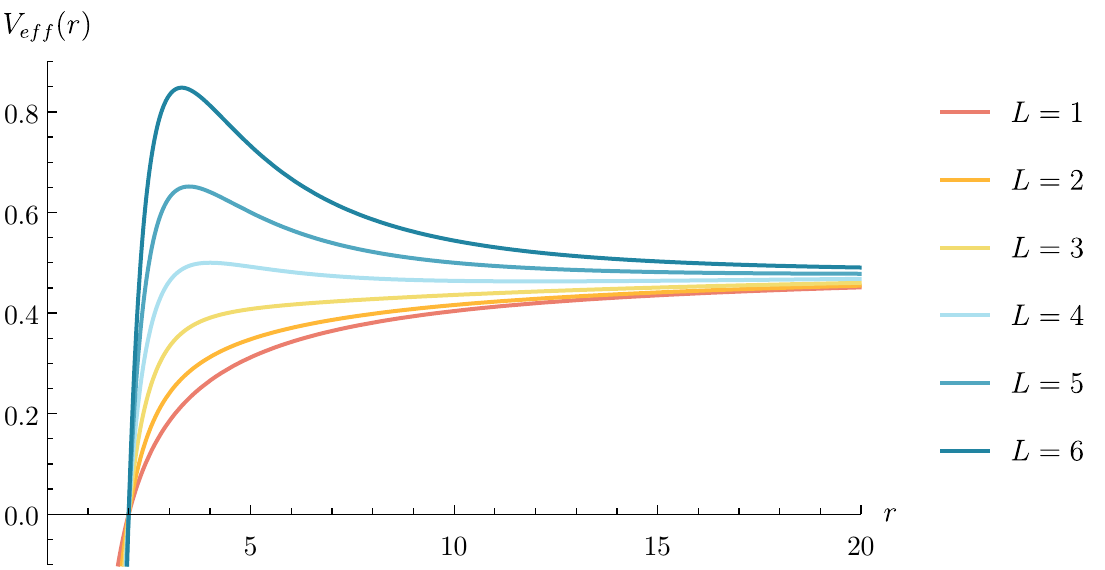}
    \captionsetup{width=.8\linewidth}
    \caption{Effective potential for massive 
    particles in General Relativity, $V_{e\!f\!f}(r)$, 
    for different values of the angular momentum $L$ ($M=1$). 
    When $L^2<12M^2$ there are no extreme points. At $L^2=12M^2$, 
    a single extremum (a saddle point) emerges. As $L$ increases, 
    two extrema appear, a minimum $r_+$ and a maximum $r_-$, 
    which gets more separated when $L$ grows. For 
    $12M^2< L^2\leq 16M^2$, the effective potential at the 
    maximum $V_{e\!f\!f}(r_-)$ remains smaller or equal to 
    the asymptotic limit, while it exceeds the latter for 
    $L^2>16M^2$.}\label{fig:GRmassive}
\end{figure}

\begin{figure}[!htb]
    \centering
    \includegraphics[width=0.7\textwidth]{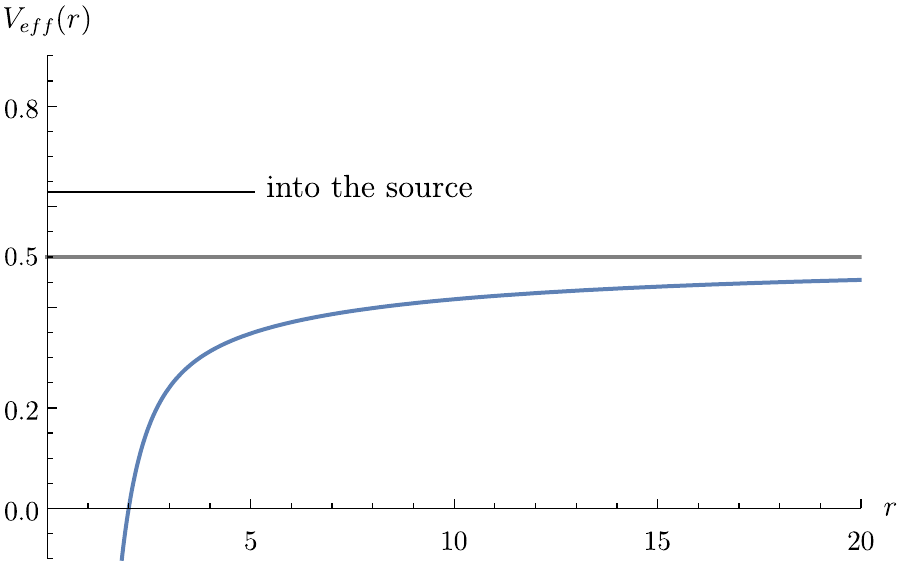}
    \captionsetup{width=.8\linewidth}
    \caption{Effective potential for massive particles 
    in General Relativity as a function of the coordinate 
    radius $r$, for $L=1.07$ (with $M=1$). In this case, 
    there are no extrema.}\label{fig:GR1massive}
\end{figure}

\begin{figure}[!htb]
    \centering
    \includegraphics[width=0.7\textwidth]{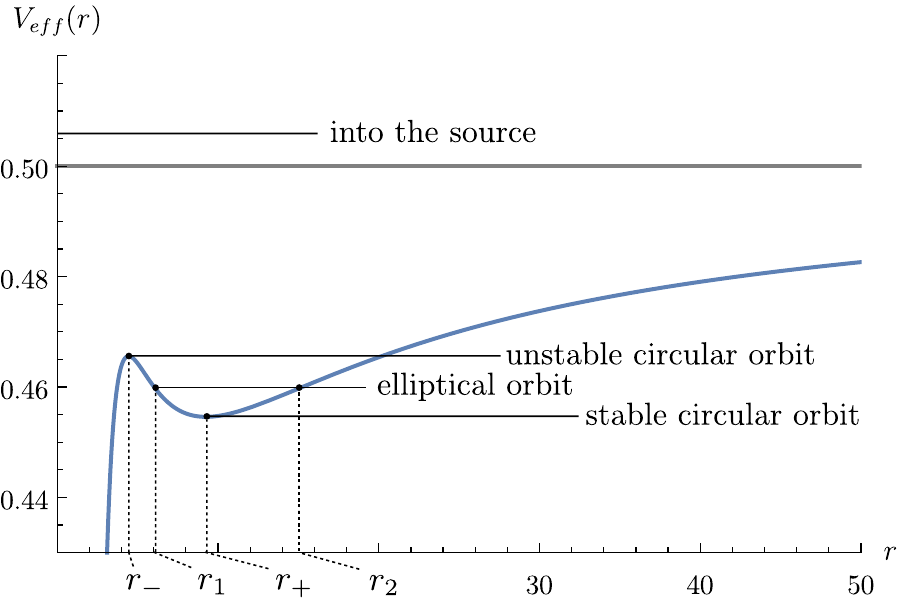}
    \captionsetup{width=.8\linewidth}
    \caption{Same as Fig.~\ref{fig:GR1massive} 
    but for $L=3.7$. Horizontal lines identify massive 
    particles with different energy states.
    Here, the potential shows both a minimum $r_+$ and a 
    maximum $r_-$, which correspond to a stable and 
    unstable circular orbit, respectively. 
    Elliptical orbits take place between 
    $r_1$ and $r_2$.}
    \label{fig:GR2massive}
\end{figure}

\begin{figure}[!htb]
    \centering
    \includegraphics[width=0.7\textwidth]{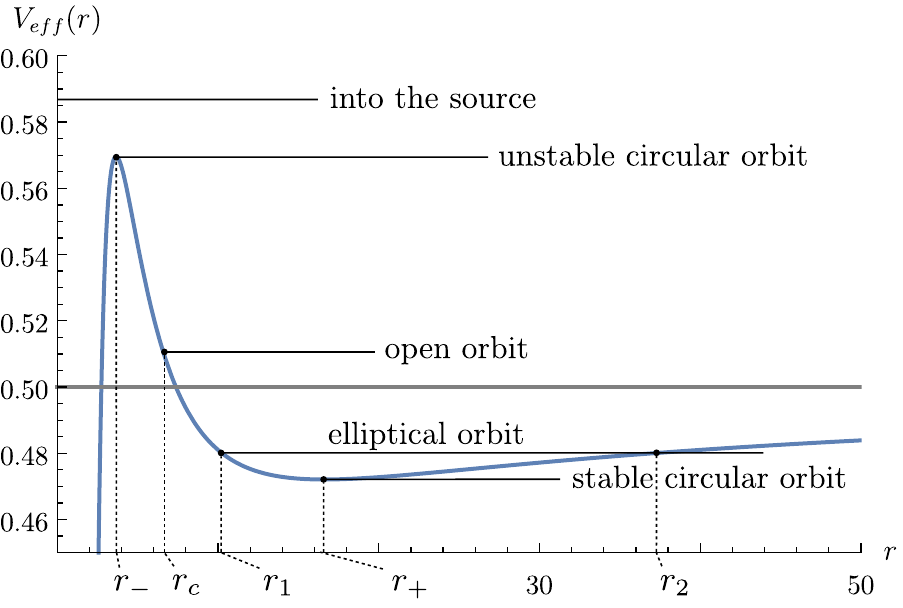}
    \captionsetup{width=.8\linewidth}
    \caption{Same as Fig.~\ref{fig:GR2massive} 
    but assuming for the angular momentum 
    of the massive particle $L=4.5$. Note that, 
    unlike the configuration shown in 
    Fig.~\ref{fig:GR2massive}, the maximum of 
    the potential is above the the asymptotic 
    value, allowing for open orbits.}
    \label{fig:GR3massive}
\end{figure}

\clearpage
\paragraph{Massive particles in General Relativity.} The 
effective potential for massive particles in GR involves 
all four terms in Eq.~\eqref{Veff}. For $\gamma=1$, the 
behavior depends on the choice of the angular momentum, 
as shown in Fig.~\ref{fig:GRmassive}. For $L^2<12M^2$, 
the potential only has imaginary roots, \textit{i.e.}, no extreme points, 
and the orbits is forced to move towards the source. 
When $L^2>12M^2$, the two roots in Eq.~\eqref{eq:rootsVeff} 
become real, corresponding to a maximum and a minimum. 
The former identifies unstable circular orbits, with a 
radius $3M\le r_-<6M$. Stable circular orbits are possible 
in correspondence of the minimum, with a radius $r_+\ge6M$. 
In GR, massive particles can exist on stable circular orbits up to $6M$, while inner circular orbits, up to $3M$, are inherently unstable. The critical radius marking the onset of stable trajectories, $r_\text{ISCO}=6M$, is known as the \textit{Innermost Stable Circular Orbit}.
In Figs.~\ref{fig:GR1massive}-\ref{fig:GR3massive} we 
show different possible configurations for the effective 
potential, and for different types of orbits. 
Depending on the value of the particle energy, the body can 
follow circular elliptical, radially bound or unbounded 
trajectories \cite{Abolghasem}.
\newpage
\subsection{Orbits of massless particles} \label{susec: orbits massless}
%
For massive particles, \textit{i.e.}, $\epsilon=0$, 
the effective potential \eqref{Veff} reads
\begin{equation}\label{massless potential}
    V_{e\!f\!f}(r)=\frac{L^2}{2r^2} -\gamma \frac{ML^2}{r^3}\ .
\end{equation}
The asymptotic value of the potential for $r\to +\infty$ 
is the same in the relativistic and in the Newtonian model, and 
tends to zero. When $r\to 0^+$, the behaviour is the same as 
described in Eq.~\eqref{limitbehavior}.

\begin{figure}[!htb]
    \centering
    \includegraphics[width=0.69\textwidth]{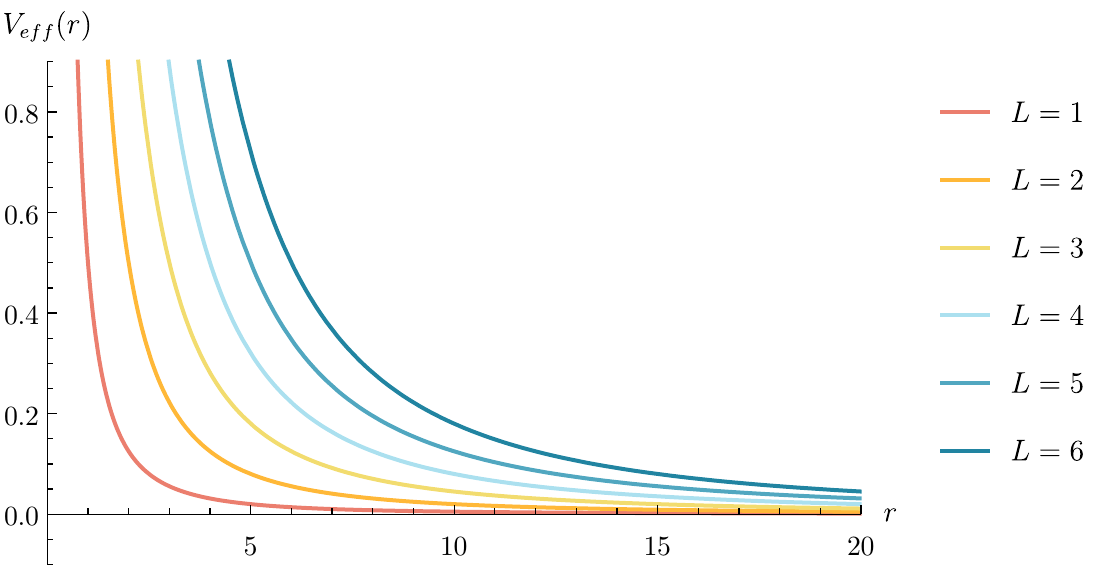}
    \captionsetup{width=.8\linewidth}
    \caption{Effective potential for massless particles 
    in Newtonian gravity as a function of $r$. In this case there are no extrema.}
    \label{fig:NGmassless}
\end{figure}

\begin{figure}[!htb]
    \centering
    \includegraphics[width=0.69\textwidth]{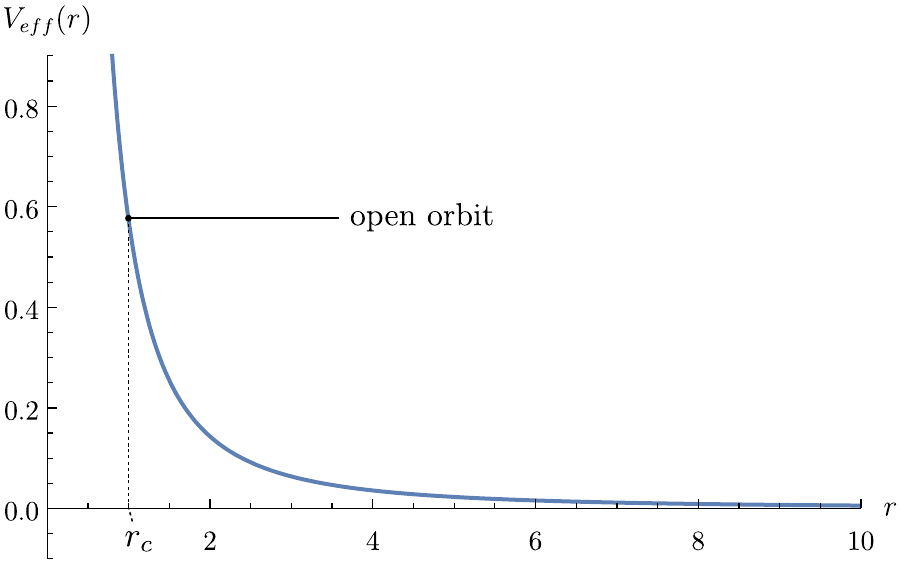}
    \captionsetup{width=.8\linewidth}
    \caption{Same as Fig.~\ref{fig:NGmassless} for a particular choice of the angular momentum $L=1.07$. In this configuration only open orbits are possible, featuring a minimum distance $r_c$.}
    \label{fig:NGmassless2}
\end{figure}

\begin{figure}[!htb]
    \centering
    \includegraphics[width=0.7\textwidth]{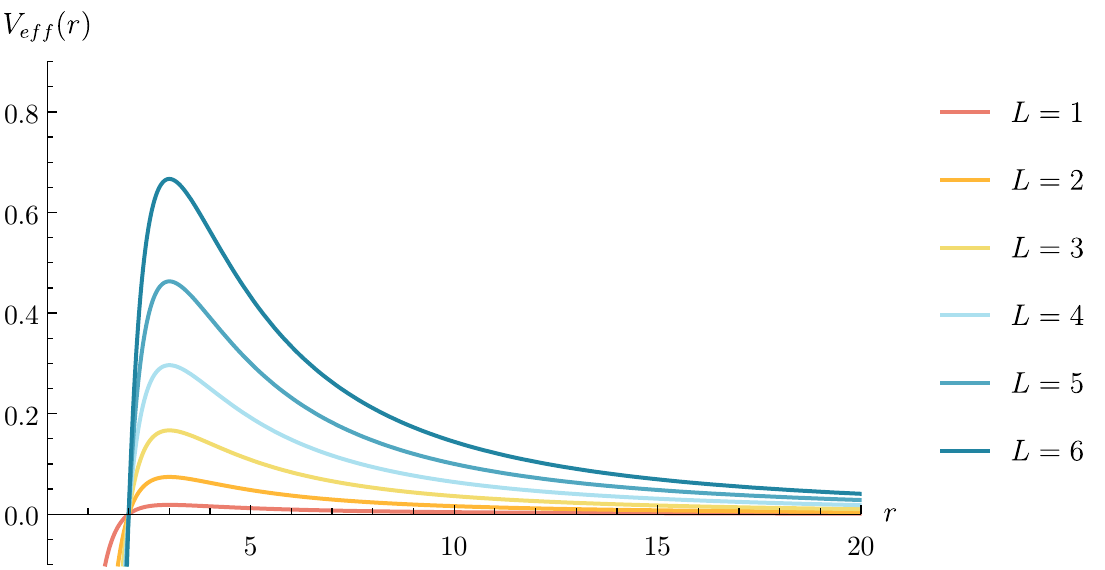}
    \captionsetup{width=.8\linewidth}
    \caption{Effective potential for massless particles 
    in General Relativity as a function of the 
    radial distance $r$ and of the angular momentum 
    $L$ (with $M=1$). The maximum of $V_{e\!f\!f}$ 
    decreases for smaller values of $L$.}
    \label{fig:GRmassless}
\end{figure}

\begin{figure}[!htb]
    \centering
    \includegraphics[width=0.7\textwidth]{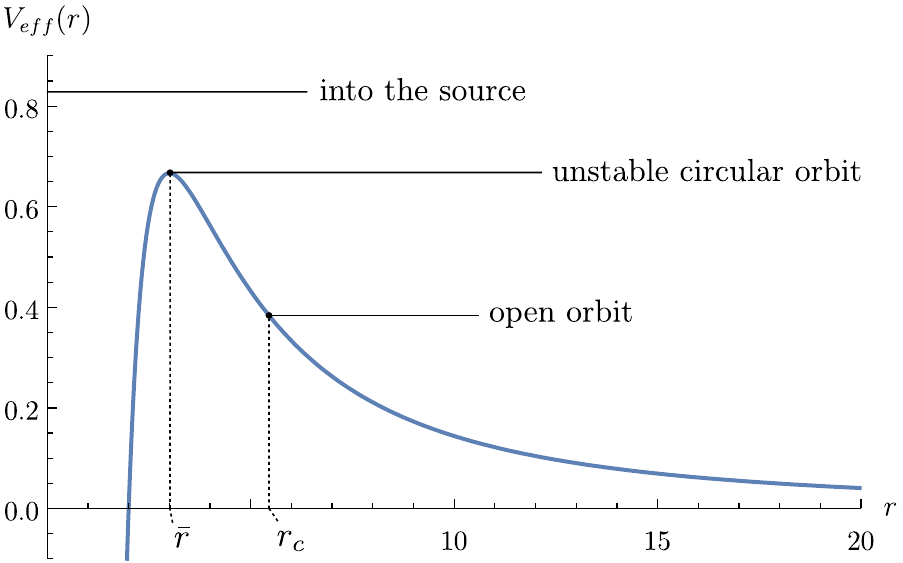}
    \captionsetup{width=.8\linewidth}
    \caption{Same as Fig.~\ref{fig:GRmassless}, but 
    assuming $L=6$.}\label{fig:GRmassless2}
\end{figure}

\paragraph{Massless particles in Newtonian gravity.} 
When $\gamma=0$, the effective potential reduces to 
$V_{e\!f\!f}(r)=\frac{L^2}{2r^2}$, which has no 
roots. Hence, for non-zero values of the angular 
momentum $L$ massless particles hit the potential 
at some distance from the source, moving 
away from it. In NG massless particles cannot 
stay on bound orbits around the source, 
and only unbounded trajectories 
are allowed (see Figs.~\ref{fig:NGmassless} and 
\ref{fig:NGmassless2}).
\newpage
\paragraph{Massless particles in General Relativity.} 
In this case, the effective potential 
\eqref{massless potential} yields a single root, 
$\bar r =3M$, which corresponds to a maximum. This 
result holds for every non-zero value of $L$, as 
shown in Fig.~\ref{fig:GRmassless} and Fig.~\ref{fig:GRmassless2}. 
At variance with the NG case, the relativistic 
description highlights three possible scenarios 
depending on the energy $\mathcal{E}$. 
Particles can indeed follow open orbits when 
$\mathcal{E}<V_{e\!f\!f}(\bar r)$. If 
$\mathcal{E}=V_{e\!f\!f}(\bar r)$, the particle 
remains on a circular orbit, although unstable. For $\mathcal{E}>V_{e\!f\!f}(\bar r)$ the 
particle falls into the source.
%
\section{Schwarzschild black holes}\label{SBH}
%
In Sec.~\ref{Schwarzschild sol}, we concluded our 
discussion on the nature of the singularities for the 
Schwarzschild metric, finding that $R_S=2M$ is 
a coordinate singularity, while $r=0$ remains a true, 
physical one. In this section, we want to study more 
in detail the spacetime region around $R_S$, adopting 
a suitable choice of coordinates. 

\begin{figure}[!htb]
    \centering
    \includegraphics[width=0.7\textwidth]{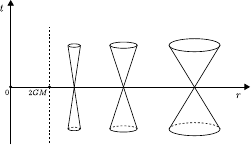}
    \captionsetup{width=.8\linewidth}
    \caption{Graphical representation in the $t$-$r$ 
    plane of light cones in the Schwarzschild metric, 
    shrinking as $r=R_S$ is approached.}
    \label{fig:lightcones}
\end{figure}

Let us consider lightlike geodesics that are radial ($\theta$ and $\phi$ both constant) within the Schwarzschild metric \eqref{Schwarzschild metric}:
\begin{equation}
    d s^2=-\left( 1-\frac{R_S}{r} \right) d t^2+\left( 1-\frac{R_S}{r} \right)^{-1} d r^2=0\ ,
\end{equation}
hence the slope of the light cones in a $t$-$r$ diagram is
\begin{equation}\label{slope}
    \frac{d t}{d r}=\pm \left( 1-\frac{R_S}{r} \right)^{-1}\ .
\end{equation}
At large distances from the source, $r\to +\infty$, 
the slope tends to $\pm 1$, as expected for a flat 
metric, since the Schwarzschild solution is asymptotically 
flat. On the other side of the domain, as we reach 
the Schwarzschild radius, $r\to R_S$, the slope diverges, $\frac{d t}{d r}\to \pm \infty$. This implies that as 
we move towards $R_S$ the slope of the light cone 
increases such that it becomes progressively narrower 
(see the sketch in Fig.~\ref{fig:lightcones}).  
Hence, an infalling particle will appear to slow down as 
it approaches $R_S$ from the perspective of external 
observers using Schwarzschild coordinates. In other words, 
we will perceive a particle taking an infinite amount of 
time to reach the Schwarzschild radius.
However such problematic description is rooted in our 
choice of coordinates, and an alternative system is 
needed. 

%
\subsection{The tortoise coordinate}
\begin{figure}[!htb]
    \centering
    \includegraphics[width=0.9\textwidth]{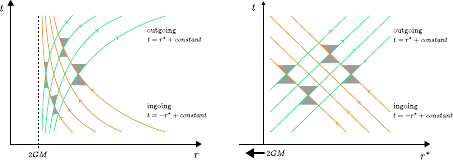}
    \captionsetup{width=.8\linewidth}
    \caption{Massless ingoing (green) and outgoing (orange) 
    geodesics of the Schwarzschild metric. (Left) Light cones 
    (in gray) deform as $r=R_S$ is approached when using the 
    Schwarzschild radial coordinate $r$. (Right) When using 
    the tortoise radial coordinate $r^*$, the Schwarzschild 
    radius is pushed to minus infinity, and the light cones 
    remain undeformed.}
    \label{fig:radial}
\end{figure}

%
A common choice involves shifting the Schwarzschild 
surface to $-\infty$ by adopting a different 
coordinate time that varies more slowly as we 
approach $R_S$. To achieve this, we introduce the 
so-called \textit{tortoise coordinate}:
\begin{equation}\label{S tortoise coordinate}
    r^*=r+2M\log \left( \frac{r}{2M} -1\right)\ ,
\end{equation}
which allows to rewrite the Schwarzschild metric 
such that the line element 
\eqref{Schwarzschild metric} reads 
\begin{equation}\label{S metric in tortoise}
    d s^2=\left( 1-\frac{2M}{r} \right) \left(-d t^2+d r^*{}^2\right)+r^2d \Omega^2\ ,
\end{equation}
where $r=r(r^*)$. The metric in these coordinates remains 
well-behaved at $R_S=2M$. Solving Eq.~\eqref{slope}, we 
find $t=\pm r^*+c$, where $c$ is an integration constant. 
The plus and minus signs correspond to outgoing and 
ingoing massless geodesics, respectively. The behavior 
of light cones in the $t$-$r$ and $t$-$r^*$ diagrams is 
illustrated in Fig.~\ref{fig:radial}, showing the 
difference in the choice of radial coordinate 
$r$ and the tortoise coordinate $r^*$.

%
\subsection{Eddington--Finkelstein coordinates}
%
We can use coordinates adapted to ingoing 
or outgoing massless particles, obtained as follows:
\begin{equation}
    v=t+r^*\ ,\qquad u=t-r^*\ .
\end{equation}
An ingoing lightlike particle is indeed characterized by $v=constant$, while an outgoing particle by $u=constant$. 
Using the Schwarzschild radial coordinate $r$, with either 
$v$ or $u$ as the coordinate time, we have the 
\textit{ingoing Eddington--Finkelstein (EF) coordinates} or 
the \textit{outgoing Eddington--Finkelstein coordinates}, respectively. In these coordinate systems, the 
Schwarzschild metric is given by
\begin{equation}
    \begin{aligned}
        \text{ingoing EF}\qquad & d s^2 =-\left( 1-\frac{2M}{r} \right) d v^2+2 d vd r +r^2d \Omega^2\ ,\\
        \text{outgoing EF}\qquad & d s^2 =-\left( 1-\frac{2M}{r} \right) d u^2-2 d ud r +r^2d \Omega^2\ .
    \end{aligned}\label{eq:EFScw}
\end{equation}
The metric shown in \eqref{eq:EFScw} is explicitly 
nonsingular, invertible, and its inverse does not 
have any divergent component. 

\begin{figure}[!htb]
    \centering
    \includegraphics[width=0.7\textwidth]{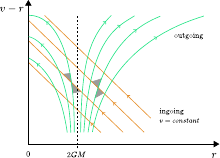}
    \captionsetup{width=.8\linewidth}
    \caption{Massless ingoing (green) and outgoing (orange) geodesics of the Schwarzschild metric in the ingoing EF coordinates. Light cones (in gray) tilt more and more as the Schwarzschild radius is approached.}
    \label{fig:EGingoing}
\end{figure}

If we consider a lightlike radial geodesic in these 
coordinates, we can compute the slope of the light 
cones in a $v$-$r$ or $u$-$r$ diagram:
\begin{equation}
    \frac{d v}{d r}=\begin{cases}
        0\ , \\
        2\left(1-\frac{2M}{r}\right)^{-1}\ ,
    \end{cases}\qquad
    \frac{d u}{d r}=\begin{cases}
        -2\left(1-\frac{2M}{r}\right)^{-1}\ , &\qquad \text{ingoing}\ ,\\
        0\ . &\qquad \text{outgoing}\ .
    \end{cases}
\end{equation}
In the ingoing EF coordinates, ingoing geodesics are straight 
lines, while outgoing geodesics are divided into two separate 
families, depending on whether $r<R_S$ or $r>R_S$, as shown 
in Fig.~\ref{fig:EGingoing}. As we move towards the 
Schwarzschild radius, the light cones tilt more and more, 
until the future light cone is completely inside $R_S$.
This reflects an important feature of the surface at 
$r=R_S=2M$, the event horizon, which is a no-return 
region. A  particle in the direction of the singularity 
$r=0$, which crosses $R_S$, cannot escape and is 
destined to fall into the source. The event horizon 
divides the spacetime into two causally disconnected 
domains: an outside observer can send signals both 
inward and outward, but can not know what happens 
inside the horizon.

\section{Perturbations of Schwarzschild black holes} \label{sec: perturbations}

Before entering into the details of relativistic perturbations 
of the Schwarzschild spacetime, it is instructive to discuss 
a scattering toy model\footnote{This example was suggested 
to the authors by Prof.~Vitor Cardoso.}, that encodes many 
properties common to the more intricate scenario we treat 
afterwards.

\subsection{A scattering toy problem}\label{sec:toymodel}

Let us consider a scattering problem defined by the following 
second-order differential equation:
\begin{equation}
\left[\frac{d^2}{dx^2}+\omega^2-2V_0\delta(x)\right]\psi(\omega,x)=i\omega\, \psi_0(x)\ ,
\label{math_ex:master}
\end{equation}
where $x\in(-\infty,\infty)$. 
Equation~\eqref{math_ex:master} features a localized 
effective potential and a source term related to the 
initial configuration\footnote{We are working in the 
frequency-domain space, where $\psi(\omega, x)$ is the 
Fourier transform of the time-domain function
$\psi(t,x)$.}. For simplicity, we also 
assume that the latter is localized, namely, 
$\psi_0(x)=\psi(t=0,x)=\delta(x-x_0)$. 
A general solution to the family of problems in Eq.~\eqref{math_ex:master} involves first solving the associated homogeneous equation and then constructing the full solution considering the source term \cite{boyce2004ede}. The homogeneous 
problem yields two solutions, identifying growing 
and decaying modes on both sides of the delta function. 
In particular, by requiring a 
purely ingoing wave as $x\rightarrow-\infty$, the solution propagating as
$x\rightarrow \infty$ is the sum of 
outgoing and ingoing modes:
\begin{equation}
\psi_\tn{L}=\begin{cases}
e^{-i\omega x}\ ,   &x\rightarrow -\infty\ ,\\
A_\tn{in}e^{-i\omega x} +A_\tn{out}e^{i\omega x}\ ,  &x\rightarrow +\infty\ .
\end{cases}
\end{equation}
Requiring continuity of the solution at $x=0$ leads to 
$A_\tn{in}+A_\tn{out}=1$. Using the field equation to compute the jump of the first derivative, we integrate the master equation \eqref{math_ex:master} within $[-\epsilon,\epsilon]$ 
as $\epsilon\rightarrow 0$:
\begin{equation}
\int_{-\epsilon}^{+\epsilon}\! dx\, \frac{d^2}{dx^2}\psi_\tn{L}
+\int_{-\epsilon}^{+\epsilon}\! dx\,\omega^2\psi_\tn{L}=
\int_{-\epsilon}^{+\epsilon}\! dx\, 2 V_0\,\delta(x)\, \psi_\tn{L} \ .
\end{equation}
The second integral of the left-hand side vanishes assuming that $\psi_\tn{L}(x)$ is 
continuous. Hence, we have
\begin{equation}
 \frac{d \psi_\tn{L}}{dx}\bigg\vert_{-\epsilon}^{+\epsilon}
=2V_0\, \psi_\tn{L}(x=0)\ .
 \end{equation}
Combining the former with the condition on the wave 
amplitude we obtain
\begin{equation}
\begin{cases}
A_\tn{in}+A_\tn{out}=1\ ,\\ 
-i\omega A_\tn{in}+i\omega A_\tn{out}+i\omega=2V_0\ ,
\end{cases}
\end{equation}
hence
\begin{equation}\label{Ain}
A_\tn{in}=1+\frac{iV_0}{\omega}\ , \qquad
A_\tn{out}=-\frac{iV_0}{\omega}\ .
\end{equation}

Let us now compute the Wronskian between $\psi_\tn{L}(x)$ 
and a second solution, labelled $\psi_\tn{R}(x)$, 
which behaves as purely outgoing at infinity, 
\textit{i.e.}, $\psi_\tn{R}(x)= e^{i\omega x}$ as 
$x\rightarrow \infty$:
\begin{equation}
    \begin{aligned}
        W&=\frac{d\psi_\tn{R}}{dx}\psi_\tn{L}-\frac{d\psi_\tn{L}}{dx}\psi_\tn{R}\\
        &=i\omega e^{i\omega x}(A_\tn{out}e^{i\omega x}+A_\tn{in}e^{-i\omega x})-e^{i\omega x}(i\omega A_\tn{out}e^{i\omega x}-i\omega A_\tn{in}e^{-i\omega x})\\
        &= i\omega A_\tn{out}e^{2i\omega x}+i\omega A_\tn{in}-i\omega A_\tn{out} e^{2i\omega x}+i\omega A_\tn{in}=2i\omega A_\tn{in} \ .
    \end{aligned}
\end{equation}
Replacing the value of $A_\tn{in}$ found before (see Eq.~\eqref{Ain}), 
we obtain an analytic expression of the Wronskian:
\begin{equation}
W=2i \omega A_\tn{in}=2i\omega +2 i^2 V_0=2i\omega -2V_0\ .
\end{equation}
The two solutions $\psi_\tn{L,R}(x)$ 
are linearly dependent if the values of $\omega$ solve the 
eigenvalue problem given by the second order equation 
\eqref{math_ex:master}, \textit{i.e.}, if they correspond to the 
QNMs of the system. In this case $W=0$, or equivalently 
$\omega=-i V_0$.

We can now focus on the inhomogeneous problem. We first 
find the general solution in Fourier space using a Green-function approach:
\begin{equation}
\psi(\omega,x)=\psi_\tn{R}\int_{-\infty}^{x}\!dx\,\frac{I\,\psi_\tn{L}}{W}+\psi_\tn{L}\int^{\infty}_{x}\!dx\,\frac{I\,\psi_\tn{R}}{W} \ ,  
\end{equation}
where $I=i\omega\,\psi_0(x)$. For $x\gg 1$, we 
can write the previous solution as
\begin{equation}
\begin{aligned}
    \psi(\omega,x)&=\psi_\tn{R}\int_{-\infty}^{\infty}\!dx\,\frac{I\,\psi_\tn{L}}{W}\\
    &=\psi_\tn{R}\int_{-\infty}^{\infty}\!dx\,i\omega \frac{\psi_\tn{L}}{W}\delta(x-x_0)\\
    &\simeq \frac{i\omega}{W}e^{i\omega x}\left[A_\tn{in}e^{-i\omega x_0}+A_\tn{out}e^{i\omega x_0}\right]\, ,\qquad x_0>0 \ .
\end{aligned}
\end{equation}
Replacing the value of $W$ in terms of $A_\tn{in}$, 
the we finally obtain 
\begin{equation}
\psi(\omega,x)=\frac{1}{2}e^{i\omega (x-x_0)}+\frac{1}{2}e^{i\omega (x+x_0)}\frac{A_\tn{out}}{A_\tn{in}}\ .
\end{equation}
We can invert the solution to find the its expression 
in the time domain:
\begin{equation}
    \begin{aligned}
        \psi(t,x)&=\frac{1}{2\pi}\int_{-\infty}^{\infty}\!dx\,e^{-i\omega t}\psi(\omega, x)\\
        &=\frac{1}{4\pi}\int_{-\infty}^{\infty}\!d\omega\,e^{i\omega[(x-x_0)-t]}+\frac{1}{4\pi}\int_{-\infty}^{\infty}\!d\omega\,\frac{A_\tn{out}}{A_\tn{in}}e^{i\omega[(x+x_0)-t]}\ ,
    \end{aligned}
\end{equation}
and, using the definition of the delta function,
\begin{equation}
    \psi(t,x)=\frac{1}{2}\delta(x-x_0-t)+\frac{1}{2\pi}\int_{-\infty}^{\infty}\!d\omega\,\frac{-iV_0}{\omega+iV_0}e^{i\omega[(x+x_0)-t]}\ . \label{math_ex:integral}
\end{equation}
The second integral has a pole at $\omega=-V_0$, corresponding to the QNM frequencies. This integral can be solved using contour 
integrals by extending the domain into the complex 
plane and applying the residue theorem. 
There are two cases to consider. 
\begin{figure}[!htb]
    \centering
    \includegraphics[width=0.45\textwidth]{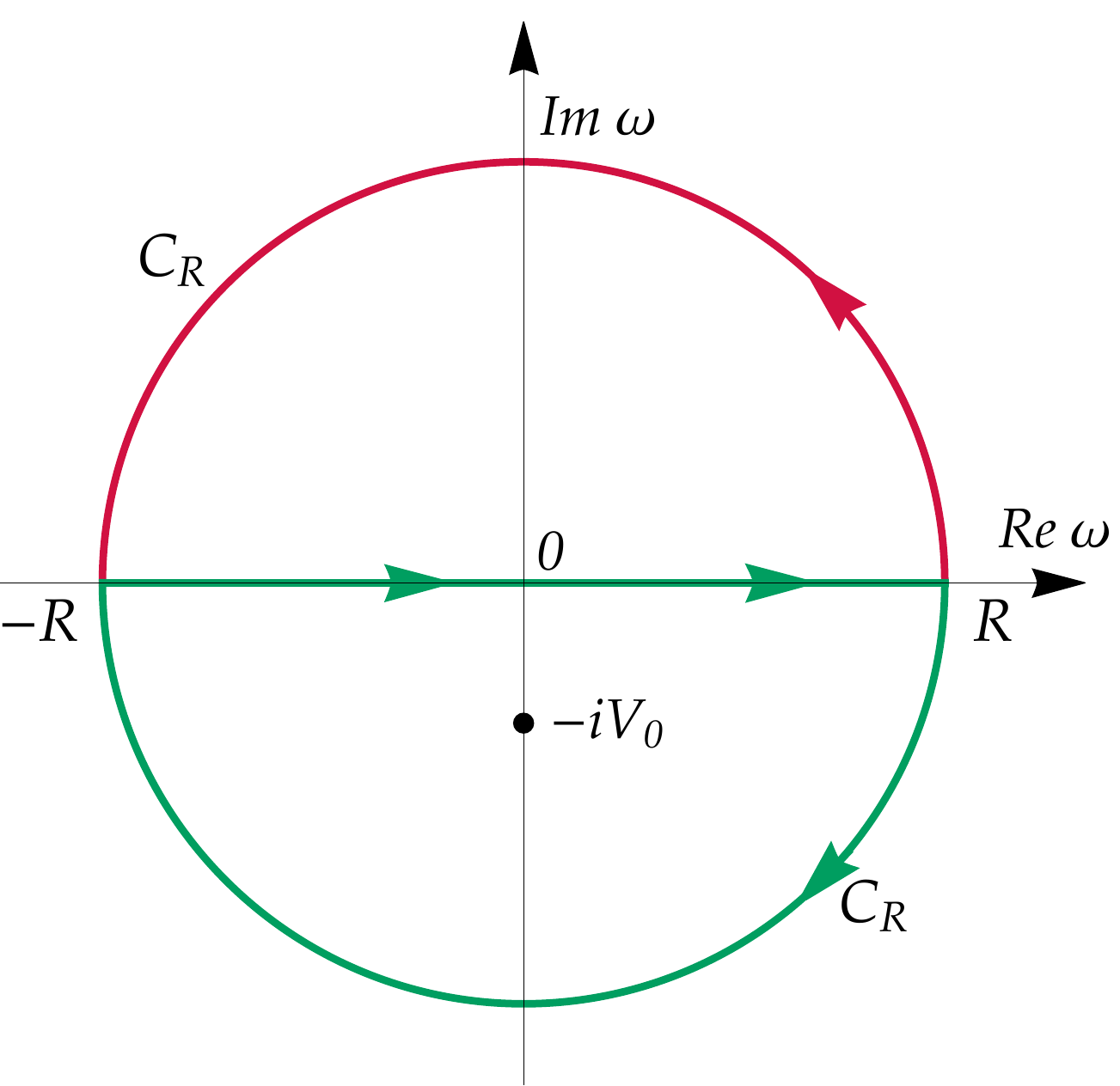}
    \captionsetup{width=.8\linewidth}
    \caption{Contour path used to perform the integral in Eq.~\eqref{math_ex:integral}. The pole corresponding to the QNM, $-iV_0$, is located in the negative imaginary panel.}\label{fig:contour}
\end{figure}
For $x+x_0-t>0$ $(t<x+x_0)$, we close the 
path in the upper panel. The integral vanishes 
as there are no poles inside the contour.
Conversely, for $x+x_0-t<0$ $(t>x+x_0)$, we close the path in the bottom panel. 
In this case, the integral over the path 
shown in Fig.~\ref{fig:contour} reads\footnote{The 
integral on the half-circle vanishes due to the 
Jordan lemma.}:
\begin{equation}
    \begin{aligned}
        \ointclockwise_\gamma\!d\omega\,f(\omega)e^{i\omega[x+x_0-t]}&=\int_{C_R}\!d\omega\,f(\omega)e^{i\omega[x+x_0-t]} + \int_{-R}^{R}\!d\omega\,f(\omega)e^{i\omega[x+x_0-t]}\\
        &=2\pi i\lim_{\omega\rightarrow -iV_0}\left[(\omega+iV_0) f(\omega)e^{i\omega[x+x_0-t]}\right]\\
        &=\frac{V_0}{2}e^{V_0(x+x_0-t)}\ ,
    \end{aligned}
\end{equation}
where $2\pi f(\omega)=-iV_0/(\omega+iV_0)$.
In summary, we obtain the following full solution:
\begin{equation}
    \begin{cases}
        \psi(t,x) = \frac{1}{2}\delta(x-x_0-t)\ , & t<x+x_0\ , \\
        \psi(t,x) = \frac{1}{2}\delta(x-x_0-t)+\frac{V_0}{2}e^{V_0(x+x_0-t)} \ , & t>x+x_0 \ .
    \end{cases}
\end{equation}
These equations provide a clear picture of the 
response of the system under a given perturbation, which can 
classified into two regimes. At early times ($t<x+x_0$), we have a \textit{prompt} 
response or a direct signal: the radiation propagates 
towards the observer without having the time to interact 
with the potential barrier and return. Later, 
for $t>x+x_0$, we also appreciate the effect due to 
the QNM. The radiation interacts with the potential 
barrier, and within a time $x-x_0+2x_0=x+x_0$, the observer 
see the trains of modes.
\begin{figure}[!htb]
    \centering
    \includegraphics[width=0.65\textwidth]{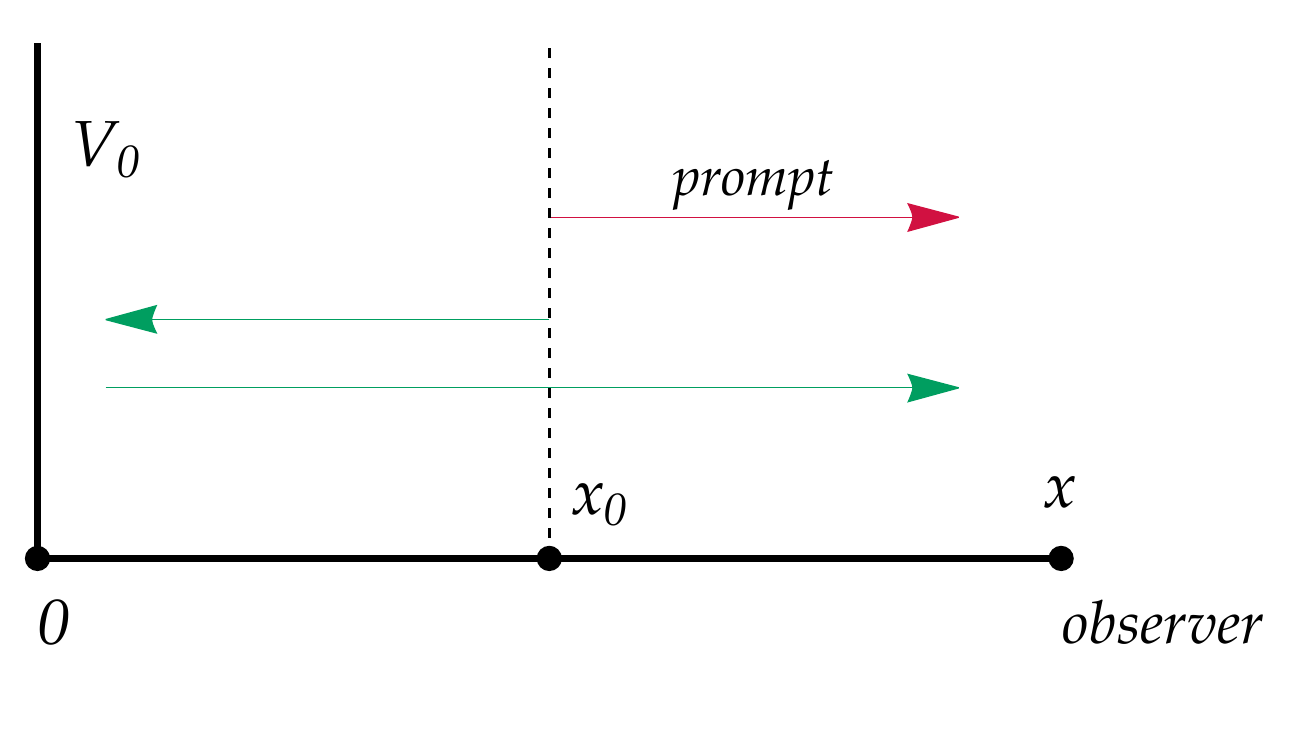}
    \captionsetup{width=.8\linewidth}
    \caption{Schematic representation of the barrier-observer 
setup for the scattering problem defined by equation \eqref{math_ex:master}. }\label{fig:barries}
\end{figure}

The picture describing the black hole response to an external perturbation remains qualitatively the same. Instead of the delta function, the scattering will be characterized by an effective potential. The time-domain response will exhibit a prompt effect originating from the initial data, while the other contribution will be absorbed by the black hole and excite its modes. After a certain interval of time, this excitation leads to the ringdown signal. Notably, the excitation of the QNMs 
is localized at a very special place: the light 
ring.
%
\subsection{Scalar field perturbations}
%
Instead of going through all the steps to compute gravitational perturbations of the Schwarzschild metric, let us focus on the perturbations induced by a massive scalar field on the background spacetime. As we will discuss at the end of this section, the results obtained for this \textit{probe} field are generic enough to be straightforwardly generalized to vector and tensor perturbations, without the need for a more complicated mathematical formalism.

We consider a scalar field $\phi$ that is 
minimally coupled with gravity, 
described by the following action:
\begin{equation}
    S=S_\tn{EH}+S_{\phi}=\int\! d^4 x\, \sqrt{-g}\,\gamma\, R + \int\! d^4x\, \sqrt{-g} \left( \kappa\, \partial_\mu \phi \,\partial^\mu \phi +\beta\, \phi^2 \right)\ .
\end{equation}
Here, $S_\tn{EH}$ and $S_\phi$ identify 
the Einstein--Hilbert and the scalar field 
actions, respectively, with $\gamma$, $\kappa$, and $\beta$ being three coupling constants. The equations of motion can be derived from the 
action using the Euler--Lagrange 
equations \cite{Lewis_2009}
\begin{equation}
    \frac{\partial \mathcal{L}}{\partial \psi}-\nabla_\mu\left(  \frac{\partial \mathcal{L}}{\partial \partial_{\mu}\psi}\right)=0\ ,
\end{equation}
where $\nabla_\mu$ represents the 
covariant derivative, (also denoted hereafter 
by a semicolon), and $\mathcal{L}(\psi, \nabla\psi)$ is 
the Lagrangian density of the system, a function of the 
fields $\psi$ and their derivatives. 
The Euler--Lagrange equations for the scalar field 
yield
\begin{equation}
    \frac{\partial \mathcal{L}_\phi}{\partial \phi}=2 \sqrt{-g} \,\beta\, \phi\ ,\qquad \frac{\partial \mathcal{L}}{\partial\partial _\mu\phi}=2 \sqrt{-g} \,\kappa\, \partial^\mu\phi\ ,
\end{equation}
which leads to
\begin{equation}\label{eom scalar}
    2 \sqrt{-g}\left( \beta\, \phi- \kappa\, \nabla_\mu\partial^\mu\phi \right)=2 \sqrt{-g}\left( \beta\, \phi- \kappa\, \Box\phi \right)=0\ .
\end{equation}
Choosing $\kappa=1$ and $\beta=\mu^2$ leads to the 
well-known Klein--Gordon equation for a scalar field with 
mass $\mu$
\begin{equation}\label{KG eq}
    (\Box -\mu^2)\phi=\frac{1}{\sqrt{-g}}\partial_\mu\left( \sqrt{-g}\, \partial^\mu \phi \right)-\mu^2\phi=0\ .
\end{equation}
The second equality can be derived from the following 
identity:
\begin{equation}
    \tensor{V}{^\mu_{;\mu}}=\frac{1}{\sqrt{-g}}\left( \sqrt{-g}\,V^\mu \right)_{,\mu}\ ,
\end{equation}
when applied to the d'Alambert operator $\Box=\nabla_\mu \nabla^\mu$ 
acting on a scalar field. Variation of the $S_\tn{EH}$ with respect 
to the metric yields the canonical Einstein Tensor, while from 
the scalar field Lagrangian
\begin{equation}
        \frac{\partial \mathcal{L}_\phi}{\partial g^{\mu\nu}} = -\frac{1}{2}g_{\mu\nu}\mathcal{L}_\phi+\sqrt{-g}\kappa\nabla_\mu\phi\nabla_\nu\phi\ ,\quad
        \frac{\partial \mathcal{L}_\phi}{\partial\partial_\sigma \tensor{g}{^\mu^\nu}}=0\ ,
\end{equation}
leading to
\begin{equation}\label{eom grav}
    R_{\mu\nu}-\frac{1}{2}g_{\mu\nu}R=\gamma^{-1}\,T_{\mu\nu}\ ,\qquad T_{\mu\nu}=\frac{1}{2}g_{\mu\nu}\left( \kappa\, \partial_\alpha \phi \partial^\alpha \phi +\beta\, \phi^2 \right)-\kappa\partial_\mu\phi\partial_\nu\phi\ ,
\end{equation}
where $T_{\mu\nu}$ is the effective stress-energy 
tensor introduced by scalar field. We have obtained, therefore, a 
set of coupled equations of motion for the scalar field 
\eqref{eom scalar} and the gravitational field \eqref{eom grav}.

Here, we assume that $\phi$ provides a small perturbation of 
the background spacetime and consider only linear-order terms 
in the scalar field. We can neglect quadratic 
contributions coming from $T_{\mu\nu}$, such that the metric and 
the scalar sector decouple. In this framework, the scalar field 
does not backreact on the metric, evolving on a fixed background given as a solution of the Einstein equations in vacuum.
%
\subsection{The master equation}\label{sec:scalpert}
%
To keep our discussion as general as possible, we consider 
here a static and spherically symmetric spacetime defined by 
Eq.~\eqref{most generic ss metric}, in which we have 
relabeled $e^{2\alpha}$ and $e^{2\beta}$ as $A$ and $B^{-1}$, 
respectively, such that the line element reads
\begin{equation}\label{ABmetric}
    d s^2=-A(r)d t^2 + B^{-1}(r) d r^2+r^2d \Omega^2\ .
\end{equation}
We can exploit the symmetries of the background and assume 
that the evolution of $\phi(t,r,\theta,\varphi)$ is 
independent of rotation, decoupling the angular 
variables $\theta$, $\varphi$ from $t$, $r$. Hence, we 
decompose the scalar field into spherical harmonics 
$Y_{\ell m}(\theta,\varphi)$: 
\begin{equation}\label{spherical harmonics}
    \phi(t,r,\theta,\varphi)=\sum\limits_{\ell=0}^\infty \sum_{m=-\ell}^\ell \frac{\psi_{\ell m}(r)}{r}e^{-i\omega t}Y_{\ell m}(\theta,\varphi)\ , \qquad Y_{\ell m}(\theta,\varphi)=N \,e^{im\varphi}P_{\ell m}(\theta)\ ,
\end{equation}
where $P_{\ell m}(\theta)$ are the Legendre polynomials 
of the second kind, $N$ is a normalization factor, 
and we have factored out the time dependence 
of the perturbation, leaving the radial function 
$\psi_{\ell m}(r)$. 
We replace Eqs.~\eqref{ABmetric} and \eqref{spherical harmonics} 
into the Klein--Gordon equation \eqref{KG eq}, to find 
a master equation for $\psi_{\ell m}(r)$. First, the 
term within round brackets in the right-hand side of 
Eq.~\eqref{KG eq} becomes
\begin{equation}\label{kg part 1}
    \begin{aligned}
        \sqrt{-g} \,\partial^\mu \phi&=\sqrt{-g}\, g^{\mu\nu}\partial_\nu \phi\\
        &=\sqrt{\frac{A}{B}}\, r^2\sin\theta \left( -A^{-1}\delta^\mu_t\partial_t + B \delta^\mu_r \partial_r +r^{-2} \delta^\mu_\theta\partial_\theta+r^{-2}\sin^{-2}\theta\,\delta^\mu_\varphi\partial_\varphi\right)\phi\ ,
    \end{aligned}
\end{equation}
where $g^{\mu\nu}=\operatorname{diag}(-A^{-1},B,r^{-2},r^{-2}\sin^{-2}\theta)$, and 
$\sqrt{-g}=\sqrt{A/B} r^2\sin\theta$. The partial 
derivatives of the scalar field expanded in spherical 
harmonics read
\begin{align}\label{derivatives scalar field}
        \partial_t \phi  = -i\omega\, \phi\quad\ ,&\quad
        \partial_r \phi  =\sum_{\ell m}\frac{e^{-i\omega t}Y_{\ell m}}{r^2}(r\psi'_{\ell m}-\psi_{\ell m})\ ,\\
        \partial_\theta \phi  =\sum_{\ell m}\frac{e^{-i\omega t}\psi_{\ell m}}{r} \,\partial_\theta Y_{\ell m}\quad\ ,&\quad
        \partial_\varphi \phi  =\sum_{\ell m}\frac{e^{-i\omega t}\psi_{\ell m}}{r} \,\partial_\varphi Y_{\ell m}\ ,
\end{align}
where a prime denotes the derivative with respect to 
the radial coordinate $r$. Next, we compute the full 
expression for $\partial_\mu \left( \sqrt{-g} \,\partial^\mu \phi \right)$: 
\begin{equation}\label{intermediate eq}
    \begin{aligned}
        \partial_\mu \left( \sqrt{-g} \,\partial^\mu \phi \right)& =\sqrt{-g}\, e^{-i\omega t} \sum_{\ell m} \left[ \tfrac{\omega^2}{A}\tfrac{\psi_{\ell m}}{r} + B \left(-\tfrac{A'}{2A}\tfrac{\psi_{\ell m}}{r^2} + \tfrac{A'}{2A}\tfrac{\psi'_{\ell m}}{r} + \tfrac{\psi_{\ell m}''}{r}\right) + B' \left(-\tfrac{\psi_{\ell m}}{2 r^2} + \tfrac{\psi_{\ell m}'}{2 r}\right)\right.\\
        &\qquad  \left. + \tfrac{\psi_{\ell m}}{r^3} \left( \csc^2\theta\, \partial^2_\varphi+\cot\theta\, \partial_\theta+\partial^2_\theta \right)\right] Y_{\ell m}\ .
    \end{aligned}
\end{equation}
Note that the previous equation can be greatly simplified 
by exploiting the properties of the spherical harmonics and 
their derivatives: 
\begin{equation}
    \begin{aligned}
        \partial^2_\phi Y_{\ell m}(\theta,\varphi) & =-m^2N \,e^{im\varphi}P_{\ell m}(\theta)=-m^2 Y_{\ell m}(\theta,\varphi)\ ,\\
        \partial_\theta Y_{\ell m}(\theta,\varphi) & = N \,e^{im\varphi}\partial_\theta P_{\ell m}(\theta)\ ,
    \end{aligned}
\end{equation}
and the same for the second derivative with respect 
to $\theta$. We can now use the following identity 
of the Legendre polynomials:
\begin{equation}
    \csc\theta\, \partial_\theta\left(\sin\theta\,\partial_\theta P_{\ell m}\right)-m^2\csc^2\theta P_{\ell m}=-\ell(\ell+1) P_{\ell m}\ ,
\end{equation}
hence
\begin{equation}
    \left(\cot\theta\, \partial_\theta + \partial^2_\theta\right)P_{\ell m}=\left(-\ell(\ell+1)+m^2 \csc^2\theta\right) P_{\ell m}\ .
\end{equation}
Therefore, the terms between parentheses in the second line 
of Eq.~\eqref{intermediate eq} can be recast simply as
\begin{equation}
    \left( \csc^2\theta\, \partial^2_\varphi+\cot\theta\, \partial_\theta+\partial^2_\theta \right)Y_{\ell m} =-\ell(\ell+1) Y_{\ell m}\ .
\end{equation}
We can therefore rewrite the Klein--Gordon equation \eqref{KG eq} as
\begin{equation}
    e^{-i\omega t} \sum_{\ell m} \left[ \tfrac{\omega^2}{A}\tfrac{\psi_{\ell m}}{r} + B \!\left(-\tfrac{A'}{2A}\tfrac{\psi_{\ell m}}{r^2} + \tfrac{A'}{2A}\tfrac{\psi'_{\ell m}}{r} + \tfrac{\psi_{\ell m}''}{r}\right) + B'\!\! \left(-\tfrac{\psi_{\ell m}}{2 r^2} + \tfrac{\psi_{\ell m}'}{2 r}\right) - \tfrac{\psi_{\ell m}}{r^3} \ell(\ell+1)  \right]\! Y_{\ell m}=0\ ,
\end{equation}
and, after further simplifications,
\begin{equation}\label{intermediate master}
    AB\frac{d^2 \psi_{\ell m}}{d r^2} + \frac{(AB)'}{2} \frac{d \psi_{\ell m}}{d r} + \left( \omega^2-\mu^2A-\frac{\ell(\ell +1)}{r^2} A- \frac{(AB)'}{2r} \right)\psi_{\ell m}=0\ ,
\end{equation}
where we have made the sum over the multipolar indices $(\ell,m)$ 
implicit. We introduce now the generalized 
tortoise coordinate $r^*$, defined by
\begin{equation}\label{tortoise r relation}
    d r^*{}^2=(AB)^{-1}d r^2\ ,
\end{equation}
which reproduces the well-known Schwarzschild tortoise 
coordinate \eqref{S tortoise coordinate} for 
$A=B=1-R_S/r$. By this change, the master equation 
\eqref{intermediate master} takes 
the particularly simple form
\begin{equation}\label{master eq}
    \frac{d^2 \psi_{\ell m}(r) }{d r^*{}^2}+\left[\omega^2-V(r)\right]\psi_{\ell m}(r)=0\ , \quad V(r)=\mu^2A+\frac{\ell(\ell +1)}{r^2} A+ \frac{(AB)'}{2r}\ .
\end{equation}
The calculations carried out so far show that the 
scalar field perturbations in a fixed, spherically-symmetric 
background are controlled by a single master equation 
\eqref{master eq}, which involves only the radial component 
of the perturbation, $\psi_{\ell m}(r)$, and resembles the Schr\"odinger equation with a scattering 
potential $V(r).$ This analogy allows for quick physical 
insights into the general features of the scattering process and enables the utilization of various resolution techniques developed in the context of quantum mechanics \cite{Berti:2014bla}.
%
\subsection{Properties of the master equation}
\begin{figure}[!htb]
    \centering
    \includegraphics[width=0.48\textwidth]{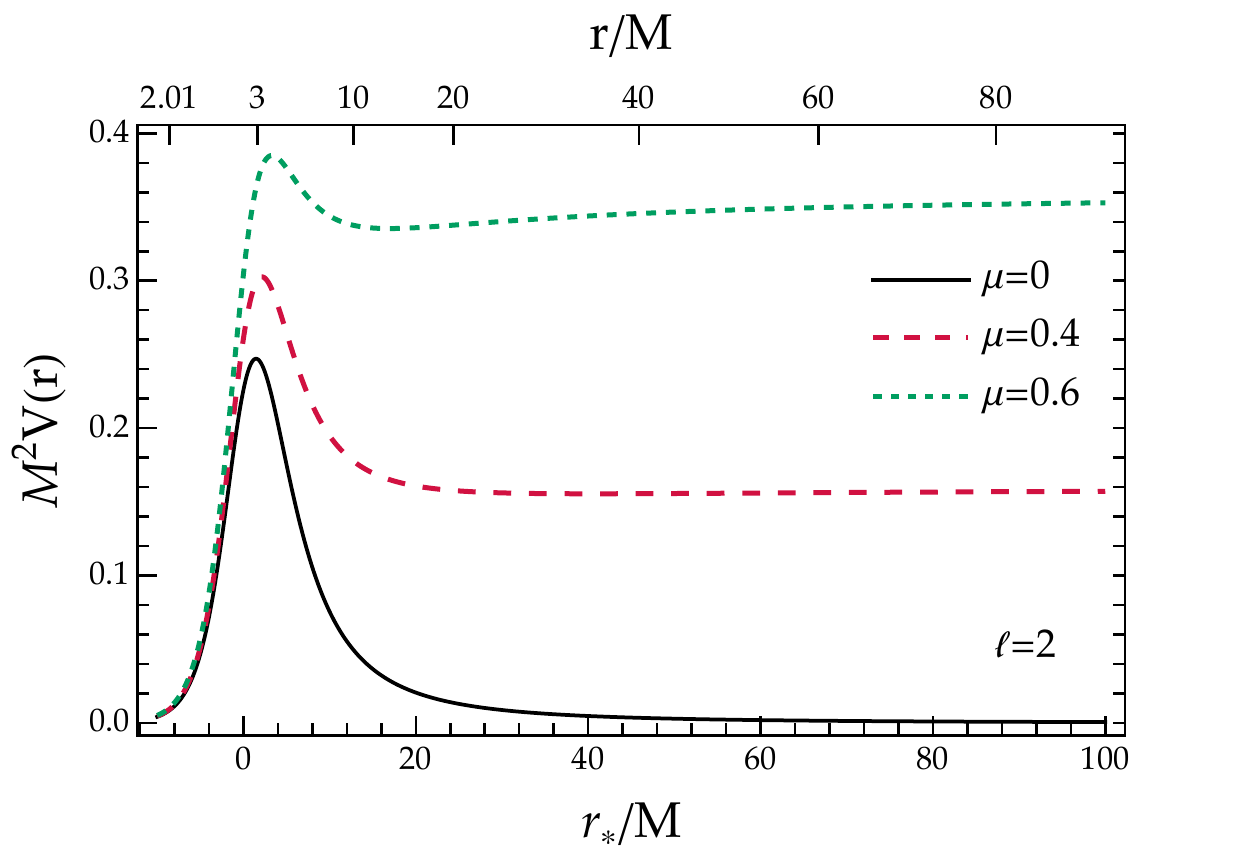}
    \includegraphics[width=0.48\textwidth]{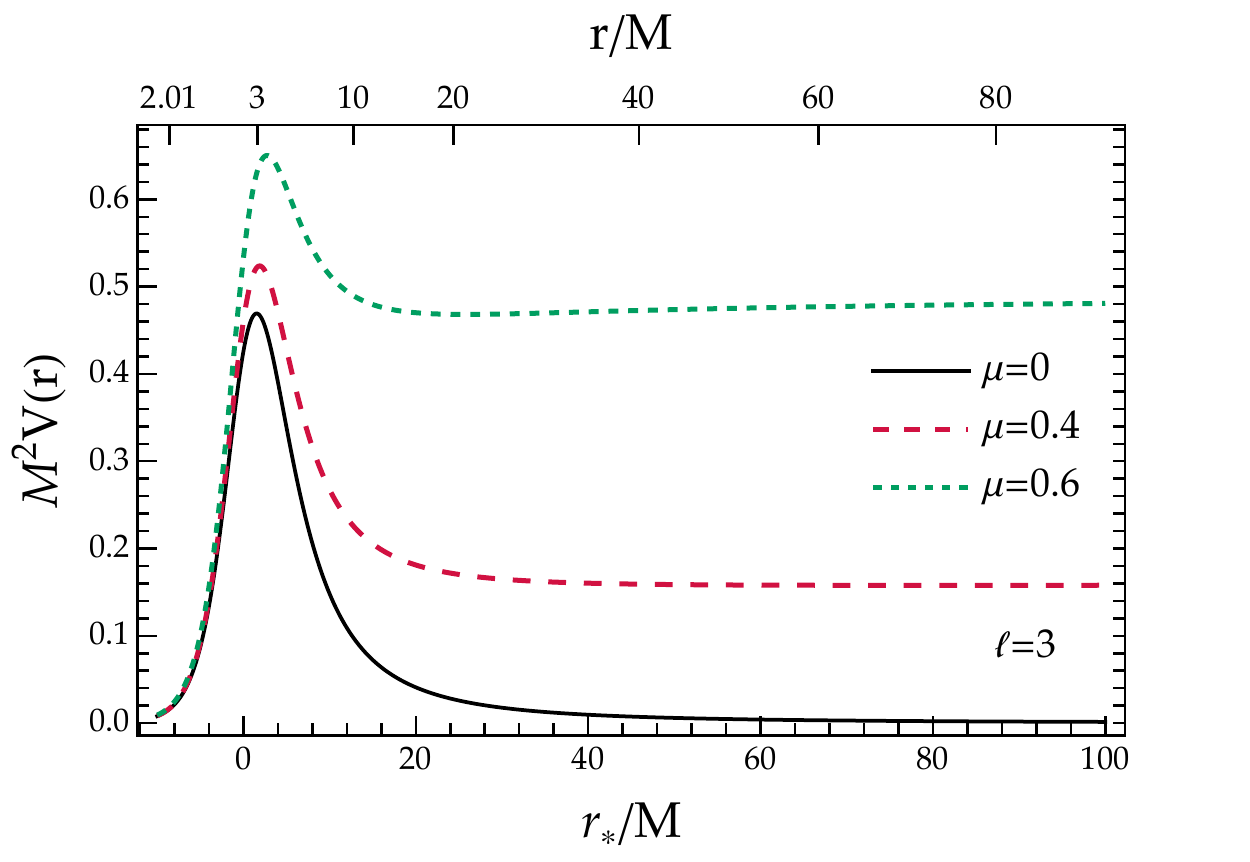}
    \captionsetup{width=.8\linewidth}
    \caption{Scattering potential of the 
    master equation \eqref{master eq} for a Schwarzschild background, with $\ell=2$ (left) and $\ell=3$ (right). Colored curves correspond to different values of 
    the scalar mass $\mu$. The top and bottom axes 
    represent values of the Schwarzschild coordinate 
    radius $r$ and the tortoise coordinate $r^*$, 
    respectively.}
    \label{fig:scatteringpot}
\end{figure}

%
We can now proceed by assuming the Schwarzschild metric, 
such that the scattering potential in Eq.~\eqref{master eq} 
reads
\begin{equation}
    V(r)=\left(1-\frac{2M}{r}\right)\left( \mu^2+\frac{\ell(\ell+1)}{r^2}+\frac{2M}{r^3} \right)\ .
\end{equation}
For a given scalar field mass $\mu$ and multipole $\ell$, 
the potential is function of the radial coordinate and: 
(i) it vanishes at spatial infinity, $r\rightarrow\infty$, (ii) 
it tends to $\mu^2$ as $r\rightarrow 2M$ (or $r^*\rightarrow -\infty)$.
Figure \ref{fig:scatteringpot} shows the behavior of $V(r)$ 
for various values of $\mu$ and $\ell$. 
Hereafter we will analyze different properties of $V(r)$ 
and the master equation to introduce the 
QNMs frequencies of a Schwarzschild BH.
For now on we will focus on the massless case $\mu=0$.

\begin{figure}[!htb]
    \centering
    \includegraphics[width=0.6\textwidth]{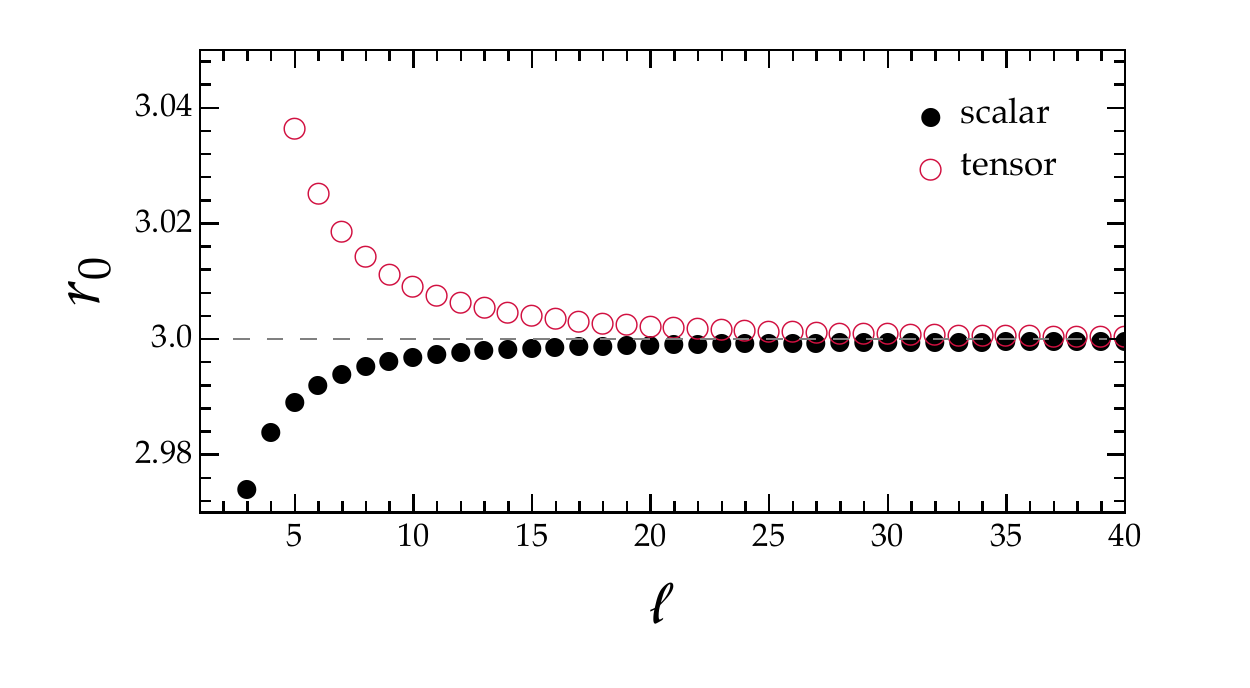}
    \captionsetup{width=.8\linewidth}
    \caption{Location of the peak of the scattering  potential \eqref{massless scattering pot} as a function of $\ell$, for scalar and tensor perturbations.}
    \label{fig:peak}
\end{figure}

\paragraph{The peak of the scattering potential.} 
Examining the two panels in Fig.~\ref{fig:scatteringpot}, 
we observe that the peak of the scattering potential 
is located \textit{suspiciously} close to 
$r=3M$, regardless of the values of  $\ell$. To 
further investigate this feature, we study the extrema 
of the scattering potential
\begin{equation}\label{massless scattering pot}
    V(r)=\left(1-\frac{2M}{r}\right)\left( \frac{\ell(\ell+1)}{r^2}+\frac{2M}{r^3} \right)\ .
\end{equation}
The critical points of $\omega^2-V(r)$ are given 
by the roots of
\begin{equation}
V'(r)=-\frac{4R_S^2}{r^5}-\frac{3R_S[\ell(\ell+1)-1]}{r^4}+\frac{2\ell(\ell+1)}{r^3}\ ,
\end{equation}
which can be found analytically, and read
\begin{equation}
    r_{\pm}=\frac{M}{2\ell(\ell+1)}\left[ 3(\ell(\ell+1)-1)\pm \sqrt{9 + \ell (\ell+1) (14 + 9 \ell (\ell+1))}\right]\ .\label{eq:peakpotential}
\end{equation}
The physical solutions, which provide 
positive radii, are given by $r_+$. 
Interestingly, as $\ell\to+\infty$, in the so called  \textit{eikonal} limit, $r_+$ tends to $3M$, as shown in Fig.~\ref{fig:peak}.
In Sec.~\ref{susec: orbits massless}, we have seen 
that $r=3M$ has a special meaning for the 
Schwarzschild spacetime, as it corresponds to the 
location of the unstable circular orbit for 
massless particles. This analysis demonstrates the remarkable correspondence between the maximum of the scattering potential and the photon ring \cite{Berti:2014bla}.

\paragraph{General master equation.} 
Thus far, we have focused on the perturbations of a 
test scalar field on a fixed Schwarzschild BH, 
finding that they reduce to the single equation 
\eqref{master eq}. 
We could follow similar steps to compute vector and tensor perturbations, studying 
the metric response. Surprisingly, while the calculations 
would become more complicated and require different 
mathematical techniques, they would lead to 
results extremely close to Eq.~\eqref{master eq}.
We can introduce a 
\textit{generalized master equation}:
\begin{equation}
    \label{general master eq}
    \begin{gathered}
        \frac{d^2 \psi_{\ell m}(r) }{d r^*{}^2}+\left[\omega^2-V_s(r)\right]\psi_{\ell m}(r)=0\ ,\qquad\\
    V_s(r)=\left(1-\frac{2M}{r}\right)\left[ \frac{\ell(\ell+1)}{r^2}+\frac{2M(1-s^2)}{r^3} \right]\ ,
    \end{gathered}
\end{equation}
where the parameter $s$ identifies the type of 
perturbation, taking values $s=0,1,2$ for scalar, 
vector, and tensor modes, respectively. For $s=0$, we indeed recover the scattering potential 
for massless scalar perturbations in the 
Schwarzschild background, as given 
in Eq.~\eqref{massless scattering pot}.

\paragraph{Boundary conditions.} 
A crucial aspect to investigate in 
Eq.~\eqref{general master eq} is the asymptotic behavior 
of the perturbations at infinity and at the horizon, 
\textit{i.e.}, the boundaries of our domain. At the event 
horizon, the potential vanishes and the solutions of the 
master equations take the form of plane waves:
\begin{equation}
    \frac{d^2 \psi_{\ell m}}{d r^*{}^2}+\omega^2\psi_{\ell m}=0 \ , \qquad\Rightarrow \qquad \psi_{\ell m}\sim e^{\pm i \omega r^*}\ .\label{eq:BC}
\end{equation}
If we assume that the radiation is completely absorbed 
at the horizon and nothing comes out from it, 
the full physical solution corresponds to the purely 
\textit{ingoing} wave $\psi\sim e^{-i\omega(t+r^*)}$. 
It is important to note, especially in light of the discussion in 
the previous section, that the horizon does not play 
any special role in the perturbations, apart from serving as a boundary condition for our solution. This aspect will be further elaborated in the following section.

On the other side of the domain, as $r^*\to +\infty$, 
the metric approaches Minkowski spacetime, and 
the master equation takes again the form \eqref{eq:BC}, 
admitting two plane wave solutions. In this case, we 
assume the condition of purely outgoing wave, \textit{i.e.}, 
that there is no incoming radiation, and consequently 
$\psi\sim e^{-i\omega (t-r^*)}$.
In summary, the wave solution behaves as
\begin{align}
    \psi_{\ell m}(r^*\to -\infty)=\psi_{\ell m}(r^*\to -\infty) \sim e^{-i\omega(t+r^*)}\  ,\quad \psi_{\ell m}(r^*\to +\infty) \sim e^{-i\omega(t-r^*)}\ .
\end{align}

\paragraph{Quasi normal modes.} 
The discussion so far highlights that 
black holes are inherently dissipative systems, 
leaking energy at the horizon and at infinity in the form of 
gravitational radiation. Consequently, the system is not 
time-symmetric, and the eigenvalue problem associated 
with Eq.~\eqref{general master eq} is non-Hermitian. This non-Hermiticity leads to complex eigenvalues, known as \textit{quasi-normal mode} (QNM) frequencies
\begin{equation}
    \omega=\omega_\tn{R}-i\omega_\tn{I}\ , 
\end{equation}
where $\omega_\tn{R,I}>0$. These QNMs characterize (part of) 
the gravitational wave response of the BH due to a 
perturbation. In a realistic physical setup (analogous to the toy model studied in Sec.~\ref{sec:toymodel}), 
the signal would feature an initial transient, whose amplitude 
is dictated by the type of perturbation, followed by a 
phase dominated by damped oscillations, with frequencies 
given by the real part of the QNMs, $f=\omega_\tn{R}/(2\pi)$. 
The inverse of the imaginary part, $\tau=1/\omega_\tn{I}$, corresponds to the damping time of each mode.

\begin{figure}[!htb]
    \centering
    \includegraphics[width=1\textwidth]{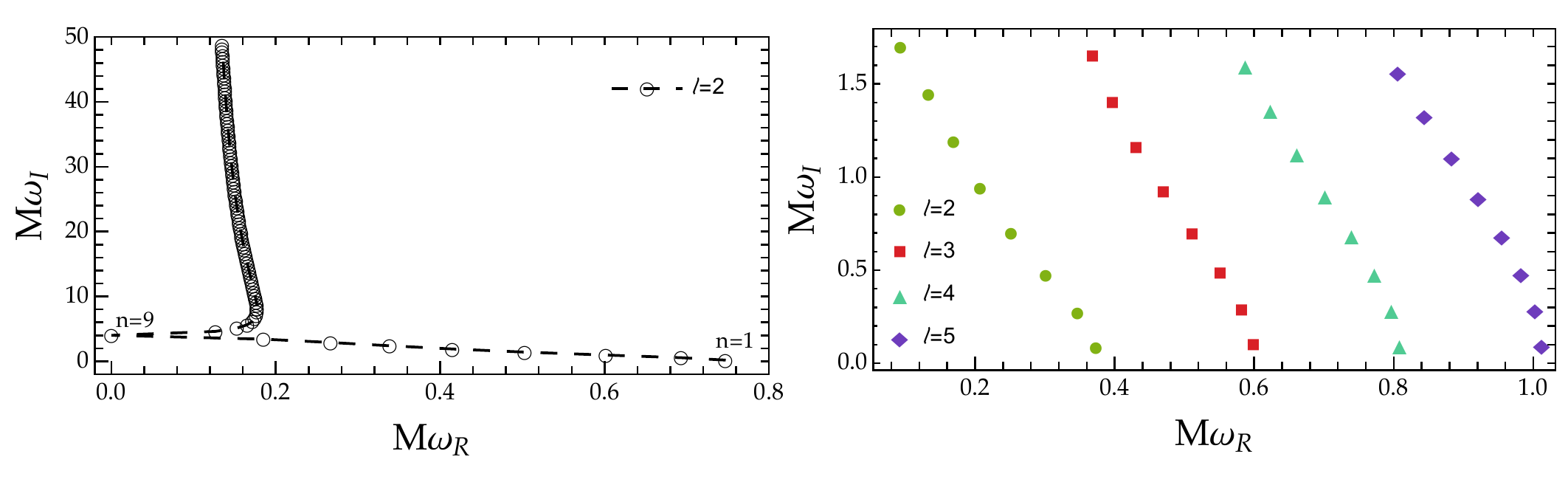}
    \captionsetup{width=.8\linewidth}
    \caption{(Left panel) 
    Real and imaginary part of the QNM for the first 100 
    overtones of the $\ell=2$ gravitational perturbations.  
    (Right panel) Same as the left panel but for $\ell=(2,3,4,5)$ 
    and considering the first 8 overtones. 
    Data from tabulated values in \cite{Berti:2005ys}.}
    \label{fig:QNMs}
\end{figure}

The left panel of Fig.~\ref{fig:QNMs} shows the real 
and imaginary part of the QNMs for $\ell=2$ 
gravitational perturbations of a Schwarzschild BH. 
Each dot corresponds to a different overtone $n$.
From the image, it is evident that $\omega_\tn{I}$ 
grows monotonically with $n$ and diverges. The 
real part has a different behavior: it decreases 
until a certain overtone and then grows, approaching 
a finite constant value. Notably, the least damped mode, 
\textit{i.e.}, with the smallest imaginary part, corresponds to $n=0$. For this reason, we expect the latter to be measured with the best accuracy. 
There is also a special mode 
with almost-vanishing real frequency, which can 
be analytically computed as 
$\omega=\pm i \ell (\ell-1)(\ell+1)(\ell+2)/6$ \cite{chandraspecial}.

Due to the complex 
nature of the frequencies, the modes diverge at both 
ends of the domain:
\begin{align}
\tn{at horizon}\quad \sim e^{-i\omega r^*}&=e^{-i\omega_\tn{R} r^*}e^{-\omega_\tn{I} r^*}\quad 
\xrightarrow[]{r^*\rightarrow-\infty}\ \infty\\
\tn{at infinity}\quad \sim e^{+i\omega r^*}&=e^{+i\omega_\tn{R} r^*}e^{+\omega_\tn{I} r^*} 
\quad \xrightarrow[]{r^*\rightarrow+\infty}\ \infty\ .
\end{align}
Hence, QNMs carry infinite energy and do not represent 
a physical state across the entire space. 
Instead, they are a localized phenomenon that, 
for a fixed $r^*$, evolves on a given time $t$. 
The larger the value of $r^*$, the larger $t$ 
must be to compensate for it. Consequently, the corresponding eigenfunctions generally do not form a complete system.
%
\subsection{Solving the master equation}
%
There are various semi-analytical and numerical 
techniques that can be utilized to solve the master 
equation \eqref{master eq}, provided appropriate boundary conditions, in order to search for the QNM 
frequencies (see \cite{Pani:2013pma} for a detailed 
review).

\paragraph{Direct integration.} 
One of the most common and efficient approaches is 
the direct integration method, which has a broader applicability beyond our specific case.
This fully numerical framework is accurate for any value of $\ell$ (and overtones) and can be extended to different physical setups, including theories of gravity beyond General Relativity and systems involving multiple coupled fields.

As discussed in the previous section, the master equation 
\eqref{master eq} yields two solutions\footnote{The 
dependence on multipolar indices is implicit here.} 
$\psi_{1,2}$, where $\psi_1$ and $\psi_2$ satisfy ingoing 
and outgoing boundary condition at the horizon and at 
infinity, respectively. The overall procedure can be 
summarized with the following steps:

\begin{enumerate}
\item Numerical Integration (Forward): Numerically integrate the master equation \eqref{master eq} 
\textit{forward} from the horizon $r_h$ to infinity $r_\infty$, 
assuming as the initial condition a purely 
ingoing solution at $r_h$, namely $\psi_1\sim e^{-i\omega r^*}$. 
In general, to improve the accuracy of our calculations, 
it is useful to compute corrections to the purely 
ingoing solution by expanding $\psi_1$ around $r_h$ as
\begin{equation}
\psi_1=\sum_{n=0}^{n_h} a_{n}(r-r_h)^{n}e^{-i\omega r^*}\ ,\label{math:BChor}
\end{equation}
where the order $n_h$ depends on the desired level of 
accuracy. The coefficients $a_n$ can be found by 
replacing Eq.~\eqref{math:BChor} into Eq.~\eqref{master eq} 
and performing a Taylor expansion around $r_h$. 
This procedure provides, order by order in $r-r_h$, 
a set of equations that can be solved for the coefficients 
$a_n$. Generally, the series of coefficients 
depends on the leading amplitude $a_0$, which can be rescaled to $a_0=1$ for our master equation.
\item Numerical Integration (Backward): integrate the master equation \eqref{master eq} 
\textit{backward}, from infinity to the horizon. In 
this case, we start the integration with an 
initial condition\footnote{The form of the solution may slightly differ from the one presented here, at both ends, when working with a different physical problem, \textit{i.e.}, master equation, as in the case of a massive field.} that is purely outgoing at 
$r_\infty$, \textit{i.e.}, $\psi_2\sim e^{i\omega r^*}$. 
Similarly to the forward integration, we can boost the accuracy of our procedure by finding the sub-leading corrections to $\psi_2$ and choosing 
\begin{equation}
\psi_2=\sum_{n=0}^{n_\infty} \frac{b_{n}}{r^n}e^{i\omega r^*}\ .
\end{equation}
The coefficients $b_n$ can be found with the same procedure 
used for Eq.~\eqref{math:BChor}, but expanding the master 
equation around $r_\infty$. Even in this case, in general, $b_n$ will depend only on the leading term $b_0$, 
which we fix to $b_0=1$. 
\item Given the two solutions above, we can build the 
Wronskian 
\begin{equation}
W(\omega)=\psi'_1\psi_2-\psi'_2\psi_1\ ,
\end{equation}
where primes denote derivatives with respect to the 
tortoise coordinate.
\item The QNMs of the systems are those for which the 
two solutions $\psi_{1,2}$ are no longer independent, 
corresponding to roots of the Wronskian. The overall 
approach translates into a findroot procedure 
to solve the equation $W(\omega)=0$.
\end{enumerate}

\paragraph{WKB.} As a second technique to compute 
QNM frequencies, we consider the WKB approach, a 
semi-analytic method that builds around the analogy 
between the master equation \eqref{general master eq} and the 
Schrodinger equation for a particle of mass $m$ and 
energy $E$, and a one-dimensional barrier 
$V(r^*)$. Here, we follow and discuss the 
calculations developed in the seminal work by 
Schulz and Will \cite{1985ApJ...291L..33S}.
We first consider a problem specified by the following 
differential equation:
\begin{equation}\label{math:wkbmaster}
\frac{d^2\psi(r^*)}{(dr^*)^2}+Q(r^*)\psi(r^*)=0\ ,
\end{equation}
which resembles our master equation with $Q=\omega^2-V_s$. 
As seen for the scalar case, the function $\psi(r^*)$ 
identifies the radial component of the full solution, which 
depends on time $(\sim e^{-i\omega t})$ and on the angular 
variables.

\begin{figure}[!htb]
\centering
\includegraphics[width=0.6\textwidth]{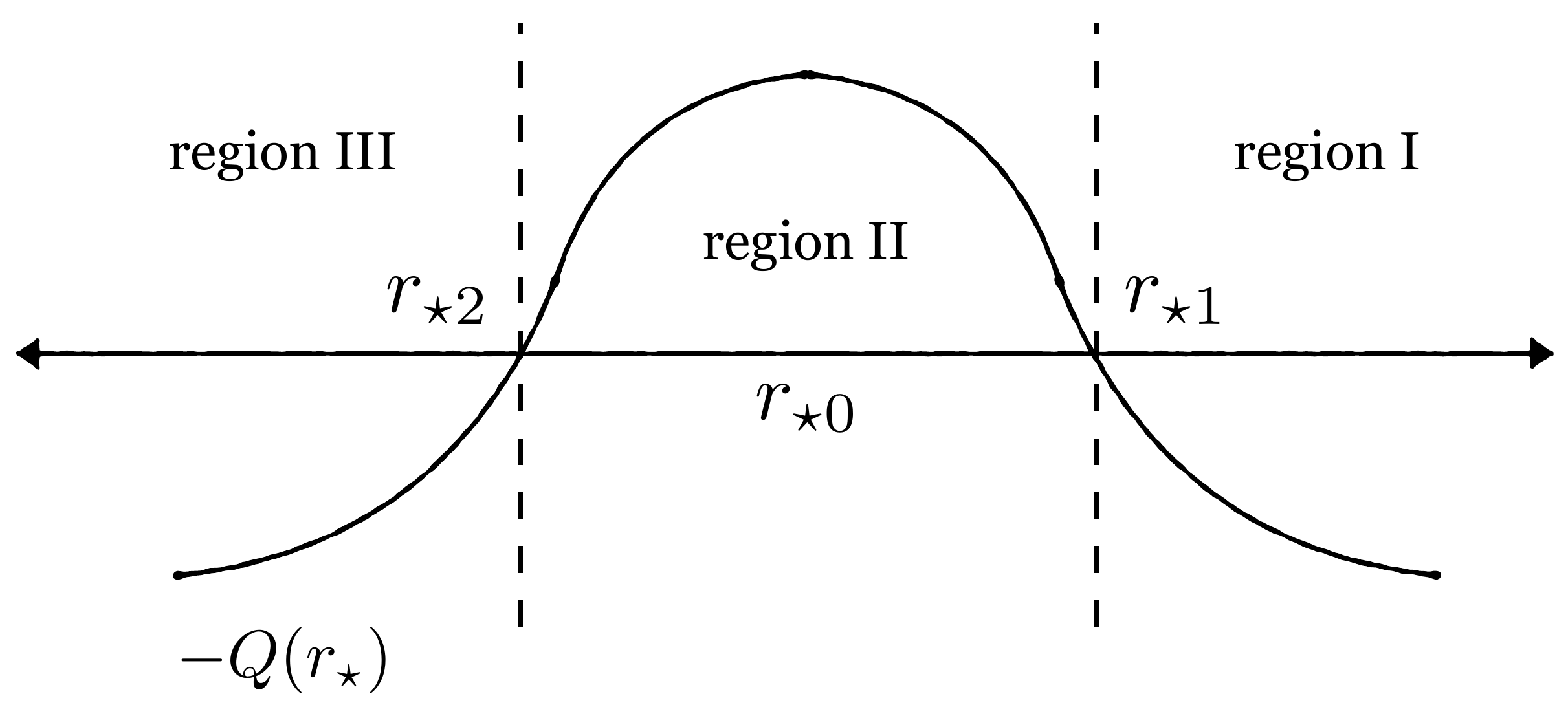}
\caption{Representation of the function $-Q(x)$. 
Adapted from \cite{1985ApJ...291L..33S}.}\label{fig:scalar_WKB}
\end{figure}
\begin{figure}[!htb]
    \centering
    \includegraphics[width=1\textwidth]{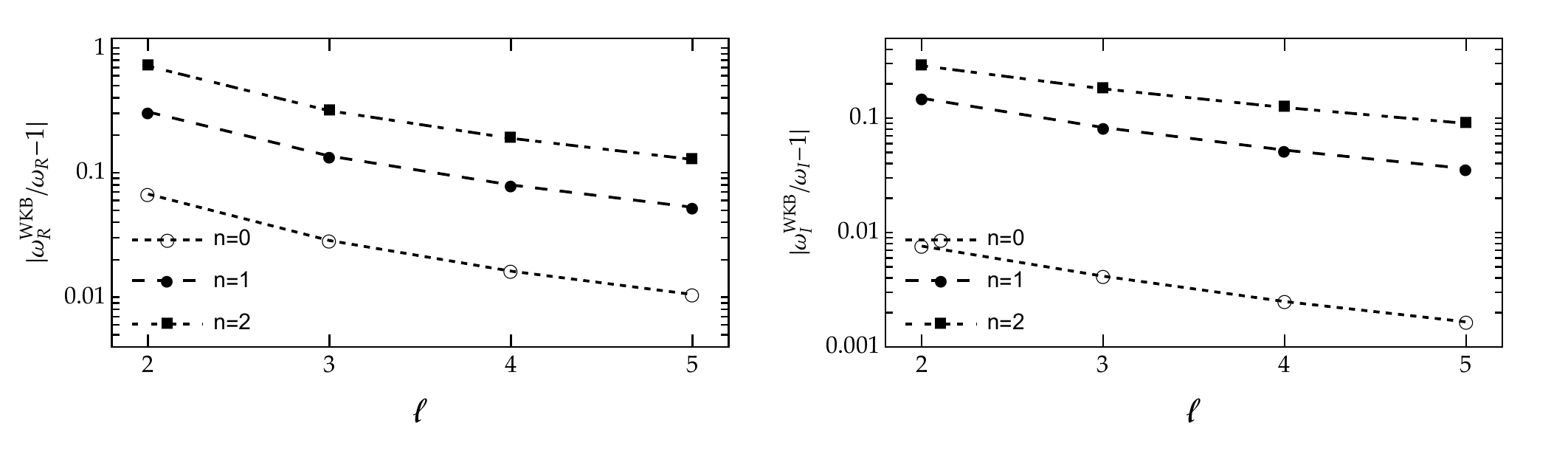}
    \captionsetup{width=.8\linewidth}
    \caption{(Left panel) 
    Relative difference between the real and imaginary part 
    of the QNM computed with the WKB approach and the values 
    obtained in \cite{Berti:2005ys}, as a function of 
    $\ell$ for the first three overtones.}
    \label{fig:WKBQNMs}
\end{figure}

The function $-Q(r^*)$ depends on the tortoise coordinate, 
has a maximum around $r^*\simeq 0$, and approaches a 
constant $Q(r^*)\rightarrow \alpha$ with $\Re(\alpha)>0$ 
at both ends of the domain, such that the solution reads
\begin{equation}
\frac{d^2\psi(r^*)}{(dr^*)^2}+\alpha\psi(r^*)=0\ , \qquad\Rightarrow\qquad \psi(r^*)\sim 
e^{\pm i\alpha r^*}\ ,\qquad \vert r^*\vert \rightarrow \infty\ ,
\end{equation}
with $e^{-i\alpha r^*}$ ($e^{i\alpha r^*}$) being outgoing 
(ingoing) modes at $+\infty$  ($-\infty$).
The basic idea behind the WKB approach is to study the 
behaviour of $Q(r^*)$ in the three zones in which 
the function is defined, finding the matching conditions 
across them. Figure \ref{fig:scalar_WKB} provides a pictorial 
representation of the regions I, II, III, and the turning 
points $r_{1}^*,r_{2}^*$, where $Q(r_{1}^*)=Q(r_{2}^*)=0$. 
In the first and third regions, the solutions for the master 
equation can be found analytically \cite{bender78:AMM}:
\begin{equation}
   Q(r^*)=\begin{cases}
       Q^{-1/4}(r^*)\exp\left[\pm i\int^{r^*}_{r_{1}^*}\!dx\,\sqrt{Q(x)}\right]\ , & \tn{region I}\ ,\\ \\
       Q^{-1/4}(r^*)\exp\left[\pm i\int^{r_{2}^*}_{r^*}\!dx\,\sqrt{Q(x)}\right]\ , & 
\tn{region III}\ .
   \end{cases} 
\end{equation}
We want now to match these solutions with region II, which is 
bounded by the turning points where $Q(r^*)=0\rightarrow V_s(r^*)=\omega^2$. 
The WKB approach works better when 
$r_{1}^*$ and $r_{2}^*$ are close, \textit{i.e.}, when 
$\vert Q(\pm \infty)\vert\gg[-Q(r^*)]_\tn{peak}$ and $V_\tn{peak}\sim\omega^2$. Following \cite{1985ApJ...291L..33S}, we approximate $Q(r^*)$ in this central zone with 
a parabola:
\begin{equation}
Q(r^*)=Q_0+\frac{1}{2}Q''_{0}(r^*-r_{0}^*)^2
+\mathcal{O}(r^*-r_{0}^*)^3\ ,
\end{equation}
with $Q_0<0$ and $Q''_0>0$. We now introduce the new variable 
$t=(4\kappa)^{1/4}e^{i\pi/4}(r^*-r_{0}^*)$, 
where $\kappa=Q_0''/2$, such that Eq.~\eqref{math:wkbmaster} can be recast in the following 
form\footnote{In this case, 
$dt=(4\kappa)^{1/4}e^{i\pi/4}dr^*$ and 
$$
Q(t)=Q_0+\frac{1}{2}\frac{2\kappa t^2}{(4\kappa)^{1/2}e^{i\pi/2}}\ .
$$}:
\begin{equation}
    \begin{gathered}
        \frac{d^2\psi(t)}{dt^2}(4\kappa)^{1/2}e^{i\pi/2}
+\left[Q_0+\frac{1}{2}\frac{2\kappa t^2}{(4\kappa)^{1/2}e^{i\pi/2}}\right]\psi(t)=0\ ,\\
\Rightarrow\quad\frac{d^2\psi(t)}{dt^2}+\left[
-\frac{iQ_0}{(2Q_0'')^{1/2}}-\frac{t^2}{4}\right]\psi(t)=0\ .
    \end{gathered}
\end{equation}
We further introduce the parameter $\nu=-\frac{1}{2}-\frac{iQ_0}{(2Q_0'')^{1/2}}$, such that 
\begin{equation}
\frac{d^2\psi}{dt^2}+\left[\nu+\frac{1}{2}-\frac{t^2}{4}\right]\psi(t)=0\ ,
\end{equation}
whose solution is given as a combination of parabolic 
cylinder functions, $D_\nu(t)$:
\begin{equation}
    \psi(t)=AD_\nu(t)+BD_{-1-\nu}(it)\ .
\end{equation}
Exploiting the asymptotic properties of these functions, 
near the horizon we find:
\begin{equation}
    \begin{aligned}
        \psi&\sim
c_1 (1-i)^\nu e^{i\pi\nu/2}\kappa^{\nu/4}(r^*-r_{0}^*)^{\nu}e^{-i\kappa^{1/2}(r^*-r_{0}^*)^2/2}\\
&\qquad+e^{-\frac{3}{4}i\pi\nu}2^{-\nu/2}\kappa^{-(1+\nu)/4}(r^*-r_{0}^*)^{-1-\nu}\left[c_2 -
c_1\frac{ie^{-i\pi\nu/2}\sqrt{2\pi}}{\Gamma(-\nu)}\right]e^{i\kappa^{1/2}(r^*-r_{0}^*)^2/2}\ ,
    \end{aligned}
\end{equation}
with $c_{1,2}$ as constants of integration, and  
$\Gamma(\nu)$ being the Euler Gamma function. The 
first and second terms of this expression identify 
ingoing and outgoing waves. 
Boundary conditions require that at the horizon, the 
outgoing modes are zero, which fixes $c_2=0$ and 
$\Gamma(-\nu)=\infty$. The latter holds if $\nu$ 
is an integer. This requirement automatically translates into a 
\textit{Born--Sommerfield quantization rule}, such that
\begin{equation}
\frac{Q_0}{\sqrt{2Q_0''}}=i\left(n+\frac{1}{2}\right)\ ,\qquad n=0,1,2,\dots\ .\label{math:scal_BSrule}
\end{equation}
The function $Q$ depends on the frequency $\omega$, 
hence Eq.~\eqref{math:scal_BSrule} turns into an algebraic 
relation which identifies a discrete set of 
complex values, the \textit{quasi-normal mode frequencies}. 
For gravitational perturbations of a Schwarzschild BH, 
$Q$ is given by Eq.~\eqref{general master eq} (with $s=2$), 
with the peak provided by Eq.~\eqref{eq:peakpotential}, allowing the computation of the values of $\omega$ for different 
$\ell$ and $n$. The WKB approximation works well for low 
overtones, \textit{i.e.}, as we have seen in Fig.~\ref{fig:QNMs} for 
modes with small imaginary parts (or large damping times), 
and for large $\ell$. The relative difference between the 
Schwarzschild QNM frequencies obtained with the WKB and the 
\textit{exact} values of \cite{Berti:2005ys} is shown 
in Fig.~\ref{fig:WKBQNMs} for $\ell=(2,3,4,5)$ and the 
first three overtones. 
The accuracy of the WKB method can be improved 
considering different approximations of the function 
$-Q(x)$ though region II and taking  into account, 
for example, higher-order approximations which go 
beyond the quadratic expansion \cite{PhysRevD.35.3632,PhysRevD.68.024018}. 

\section{Conclusions} \label{sec: conclusions}
In this notes, we have explored the key features of non-rotating 
black holes in General Relativity. Alongside the properties of the 
Schwarzschild solution, we examined in detail the motion of 
massive and massless test-bodies, along with their fundamental frequencies. 

We introduced a general framework for computing relativistic 
perturbations of spherically-symmetric and static spacetimes. Instead of elaborating on all the calculations required for tensor perturbations, 
we focused on the response induced by a test scalar field 
propagating on a fixed geometry. This approach allowed us to control perturbations using a single master equation, which can be extended to more complicated scenarios involving vector and tensor modes. Consequently, we introduced the concept of black hole oscillations. We have discussed the main properties of the Schwarzschild 
quasi-normal modes and their connection with the dynamics of massless 
particles in the background spacetime. Finally, we described two 
numerical methods, namely, the direct integration and the WKB approach, 
commonly used to compute the actual values of quasi-normal modes. 

\section*{Acknowledgements}

The authors acknowledge the contribution of the COST Action CA18108 ``Quantum gravity phenomenology in the 
multi-messenger approach''.

The work of M.~A.~has been partially supported 
by Agencia Estatal de Investigaci\'on (Spain) under grant 
PID2019-106802GB-I00/AEI/10.13039/501100011033, by the 
Regional Government of Castilla y Le\'on (Junta de Castilla y Le\'on, Spain), 
and by the Spanish Ministry of Science and Innovation MICIN and the 
European Union NextGenerationEU (PRTR C17.I1). 

\addcontentsline{toc}{section}{References}\bibliographystyle{utphys}
\bibliography{bib}

\end{document}